\documentclass[12pt]{iopart}
\usepackage[dvipdfmx]{graphicx}
\usepackage{url}

\def\regmark{{\ooalign{\hfil\raise.07ex\hbox{\tiny R}\hfil\crcr
                    {\scriptsize\mathhexbox20D}}}}
\begin{document}

\title{Computational materials design of attractive Fermion system with large negative effective $U$ in the hole-doped Delafossite of CuAlO$_2$, AgAlO$_2$ and AuAlO$_2$}

\author{A Nakanishi, T Fukushima, H Uede and H Katayama-Yoshida}
\address{Graduate School of Engineering Science, Osaka University, 1-3 Machikaneyama, Toyonaka, Osaka 560-8531, Japan}
\ead{fuku@mp.es.osaka-u.ac.jp}
\begin{abstract}
In order to realize super-high-critical temperature $(T_c)$ superconductors ($T_c > 1,000$ K) based on general design rules for negative effective $U$ $(U_{\rm eff})$ systems by controlling purely-electronic and attractive Fermion mechanisms, we perform computational materials design (CMD$^\regmark$) for the negative $U_{\rm eff}$ system in hole-doped two-dimensional (2D) Delafossite CuAlO$_2$, AgAlO$_2$ and AuAlO$_2$ from {\it ab initio} calculations.
It is found that the large negative $U_{\rm eff}$ in the hole-doped attractive Fermion systems for CuAlO$_2$ ($U_{\rm eff} = -4.53$ eV), AgAlO$_2$ ($U_{\rm eff} = -4.88$ eV), AuAlO$_2$ ($U_{\rm eff} = -4.14$ eV).
These values are 10 times larger than that in hole-doped three-dimensional (3D) CuFeS$_2$ ($U_{\rm eff} = -0.44$ eV).
For future calculations of the $T_c$ and phase diagram by quantum Monte Carlo simulations, we propose the negative $U_{\rm eff}$ Hubbard model with the anti-bonding single $\pi$-band model for CuAlO$_2$, AgAlO$_2$ and AuAlO$_2$ by using the parameters obtained from {\it ab initio} electronic structure calculations.
The behavior of $T_c$ in the 2D Delafossite of CuAlO$_2$, AgAlO$_2$ and AuAlO$_2$ and 3D Chalcopyrite of CuFeS$_2$ shows the interesting chemical trend, {\it i.e.,} $T_c$ increases exponentially ($T_c \propto \exp[-1/|U_{\rm eff}|]$) in the weak coupling regime $|U_{\rm eff} (-0.44\ {\rm eV})| < W (\sim 2\ {\rm eV})$ (where $W$ is the band width of the negative $U_{\rm eff}$ Hubbard model) for the hole-doped CuFeS$_2$, and then $T_c$ goes through a maximum when $|U_{\rm eff} (-4.88\ {\rm eV}, -4.14\ {\rm eV})| \sim W (2.8\ {\rm eV}, 3.5\ {\rm eV})$ for the hole-doped AgAlO$_2$ and AuAlO$_2$, and finally $T_c$ decreases with increasing $|U_{\rm eff}|$ in the strong coupling regime, where $|U_{\rm eff} (-4.53\ {\rm eV})| > W (1.7\ {\rm eV})$, for the hole-doped CuAlO$_2$.
In this strong coupling regime, one can expect that $T_c = 1,000 \sim 2,000$ K by assuming the relation of the very strong coupling as $2\Delta / k_{\rm B}T_c = 50 \sim100$ and the superconducting gap $\Delta \sim |U_{\rm eff}| = 4.53\ {\rm eV} \sim 50,000$ K.
\end{abstract}

\submitto{\JPCM}
\maketitle

\section{\label{sec:introduction}Introduction\protect\\}
\begin{figure}[t]
\begin{center}
\includegraphics[width=8cm,clip]{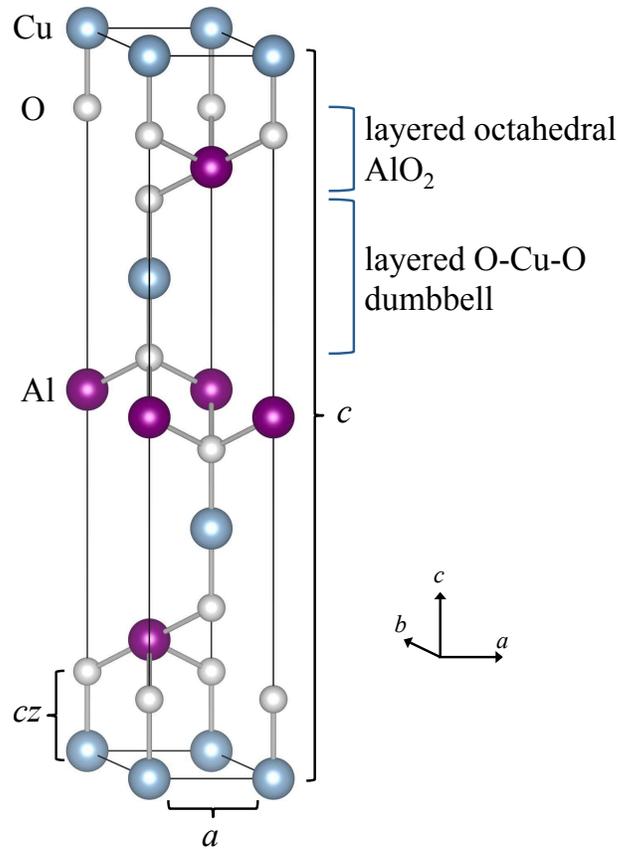}
\caption{Crystal structure of the two-dimensional (2D) Delafossite of CuAlO$_2$.
The 2D stacked layered structures of the O-Cu-O dumbbells and layered octahedral AlO$_2$ layers are formed natural super-lattices in Delafossite.
The O-Cu-O dumbbells are stacked in the 2D frustrated triangular-lattices leading to a flat-band structure (see Fig.\ \ref{band}).
Delafossite~\cite{Friedel_1} was firstly discovered by Charles Friedel ({\it French chemist and mineralogist at Sorbonne, was a student of Louis Pasteur, and great-grand father of Jacques Friedel}), and it is well known that the name of Delafossite was given in honor of Gabriel Delafosse ({\it French mineralogist and crystallographer}).}
\label{crystal}
\end{center}
\end{figure}

\begin{figure}[t]
\begin{center}
\includegraphics[height=8cm,clip]{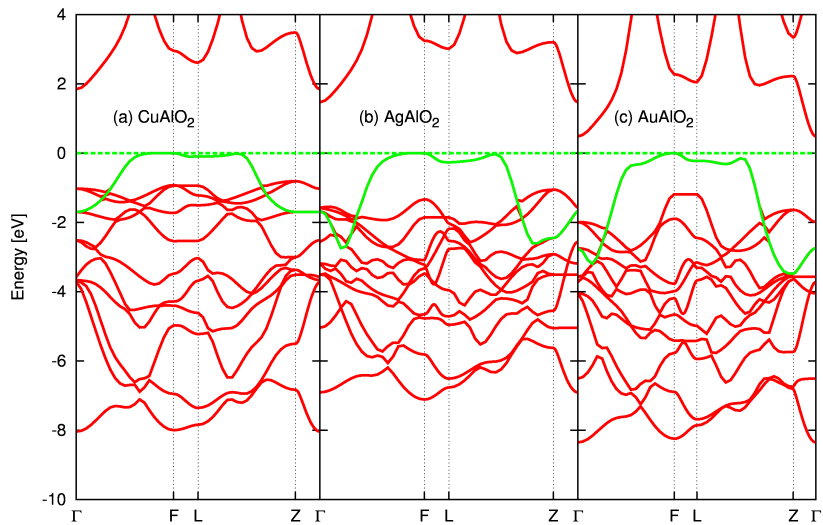}
\caption{Calculated band structures of the 2D Delafossite of (a) CuAlO$_2$, (b) AgAlO$_2$ and (c) AuAlO$_2$ by LDA.
Energy is measured from the top of the valence band maximum (VBM).
In the vicinity of VBM, the anti-bonding flat $\pi$-band (band width $W$) can be seen in the 2D frustrated triangular lattices, which is caused by the 2D stacked layered structures of the O-Cu-O dumbbell and octahedral AlO$_2$ layers in the 2D Delafossite structure (see Fig.\ \ref{crystal}).
The frustrated and very flat-band at the VBM is caused by the O$_{2p_z}$-Cu$_{3d(3z^2-r^2)}$-O$_{2p_z}$ anti-bonding $\pi$-band in the pseudo-2D frustrated triangular-lattices~\cite{HKY_2,HKY_3}.
The flat-bands are indicated by the green-color lines.
The calculated $W$ of the anti-bonding flat $\pi$-band is as follows; $W = 1.7$ eV for (a) CuAlO$_2$, $W = 2.8$ eV for (b) AgAlO$_2$, and $W = 3.5$ eV for (c) AuAlO$_2$.
The chemical trends of $W$ in the flat-band depend on the $p$-$d$ hybridization between Cu$_{3d}$, Ag$_{4d}$, Au$_{5d}$ and O$_{2p}$ orbitals.}
\label{band}
\end{center}
\end{figure}

In 1873, Charles Friedel (1832-1899, {\it French chemist and mineralogist}) discovered Delafossite~\cite{Friedel_1}, and the name of the Delafossite was given in honor of Gabriel Delafosse (1796-1878, {\it French mineralogist and crystallographer}).
The two-dimensionally (2D) stacked layered structures (natural super-lattices) of the O-Cu-O, O-Ag-O and O-Au-O dumbbells and the octahedral AlO$_2$ layers in the Delafossite of CuAlO$_2$, AgAlO$_2$ and AuAlO$_2$ (see Figs.\ \ref{crystal} and \ref{band}) show the frustrated and very flat-band structures at the valence band maximum (VBM) caused by the O$_{2p_z}$-Cu$_{3d(3z^2-r^2)}$-O$_{2p_z}$, O$_{2p_z}$-Ag$_{4d(3z^2-r^2)}$-O$_{2p_z}$, and O$_{2p_z}$-Au$_{5d(3z^2-r^2)}$-O$_{2p_z}$ anti-bonding $\pi$-bands in the pseudo-2D frustrated triangular-lattices~\cite{HKY_2,HKY_3}.

Kawazoe {\it et al.}~\cite{Kawazoe_4} discovered that the as-grown Delafossite of CuAlO$_2$ by the pulse laser deposition is a transparent $p$-type conductor (semiconductor) due to the unintentional $p$-type dopants.
Transparent $p$-type conductors are rare but necessary for the $p$-$n$ junction of transparent-oxides electronics and high-efficient photovoltaic solar-cells. 
Many industrial applications of the transparent $p$-type CuAlO$_2$ for flat-panel displays, photovoltaic solar-cells, touch panels, and high-efficient thermoelectric-power materials with $\sim$1\% hole-doping are expected~\cite{HKY_2,HKY_3,Koyanagi_5,Funashima_7}.
Hamada and Katayama-Yoshida proposed by {\it ab initio} electronic structure calculations that the unintentional $p$-type dopants are the Cu-vacancies (in the Cu-poor crystal growth condition) or Oxygen interstitials (after the post-annealing in the oxygen-rich condition) because of the small formation energies of the Cu-vacancy (single acceptor) and Oxygen interstitial (double acceptor) depending on the crystal growth conditions~\cite{Koyanagi_5,Hamada_6}. 

Based on the pseudo-2D flat-band structures of the frustrated triangular-lattices of the O-Cu-O dumbbells in CuAlO$_2$, Katayama-Yoshida {\it et al.} performed computational materials design (CMD$^\regmark$) of high-efficient thermoelectric-power materials~\cite{HKY_3,Funashima_7,HKY_11}, transparent spintronics materials by 3$d$ transition metal-doped dilute magnetic semiconductors~\cite{HKY_3,Kizaki_8,Kizaki_9,KSato_10}, the control of the pressure dependence of the electronic structures and band-gaps~\cite{Nakanishi_12}, the role of the self-interaction correction (SIC) in the quasi-particle spectrum and band-gap energy~\cite{Nakanishi_13}, and the high-$T_c$ superconductors $(T_c\sim50\ {\rm K})$ caused by the strong electron-phonon interaction in the nested 2D Fermi surfaces~\cite{HKY_2,Nakanishi_14,Nakanishi_15}.
Katayama-Yoshida {\it et al.}~\cite{HKY_2,HKY_3} suggested a new application of CuAlO$_2$ for transparent superconductors~\cite{HKY_2} and high-efficient thermoelectric-power materials with the large Seebeck coefficient and vacancy-induced small phonon thermal conductivity~\cite{HKY_3,Funashima_7,HKY_11} using the flat-band structures and small Cu-vacancy formation energy.
Katayama-Yoshida {\it et al.}~\cite{HKY_2,HKY_3} calculated the Fermi surface of the $p$-type doped CuAlO$_2$ by using the rigid band model and Full-potential Linearized Augmented Plane Wave (FLAPW) method, and proposed that the 2D nested Fermi surface may cause a strong electron-phonon interaction and a transparent superconductivity for visible light due to the large band gap ($\sim$3.0 eV).
However, the calculation of $T_c$ based on {\it ab initio} electronic and phonon calculations was not carried out in that time.
In 2012, Nakanishi and Katayama-Yoshida~\cite{Nakanishi_14} reported by {\it ab initio} calculations that it is possible to realize the $T_c$ of around 50 K due to the very strong electron-phonon interaction $\lambda$ $(\lambda \sim 0.931$ and $T_c \sim 50$ K for CuAlO$_2$ in the case of the optimized hole-doping~\cite{Nakanishi_14}, $\lambda \sim 0.9$ and $T_c \sim 40$ K for AgAlO$_2$ and $\lambda \sim 0.45$ and $T_c \sim 3$ K for AuAlO$_2$~\cite{Nakanishi_15}) by the 2D nested Fermi surfaces originating from the pseudo-2D flat-band structures of the O-Cu-O dumbbells, if we can do an ideal hole-doping without disturbing the flat-band structures in CuAlO$_2$.
The $T_c$ of CuAlO$_2$ estimated by {\it ab initio} calculations is the highest one by electron-phonon interaction in the normal pressures up to now~\cite{Nakanishi_14,Nakanishi_15}. 

We need to develop more knowledge-based CMD$^\regmark$ and searching methodologies for the discovery of new materials with wide visibility, less labor-intensive and theory-based efficiency.
The most of the experimental efforts to find new high-$T_c$ superconductors are based on the 20-th Century's brute-force and trial-and-error method (with narrow-visibility, labor-intensive and low success-rate) or accidental discovery method (with no-excellence, no-leadership and no-continuous innovations).
The new methodology to design and find new functional materials based on the quantum theory and knowledge-based digital-data bases is one of the most important issues for the development of CMD$^\regmark$ to realize the super-high-$T_c$ superconductors for the real industrial applications.
The social requirements of the knowledge-based CMD$^\regmark$ for the post-industrialized and knowledge-based society are stimulated by the change of the industrial structures and hierarchy from the industrialized society to the knowledge-based society.
In the knowledge-based societies, it is very difficult to keep and continue the 20-th Century's brute-force and trial-and-error method or accidental discovery method for the fabrication of new functional materials due to the labor shortage, high-costs and low success-rate.

Our grand challenge of the knowledge-based CMD$^\regmark$ is the design of super-high-$T_c$ superconductors $(T_c>1,000\ {\rm K})$ with a $s$-wave superconducting gap ($\Delta$), and the CMD$^\regmark$-guided realization of the super-high-$T_c$ superconductors made from the environment-friendly and non-critical or non-strategic elements, such as Cu, Fe, Al, Mg, Si, Zn, Sn, Ca, O, N, {\it etc}.
For the real industrial applications of the superconductivity, such as electric-power transportations, electric-power storages, magnets and motors, operating at the room temperature ($\sim$300 K), we do need super-high-$T_c$ superconductors with the $s$-wave $\Delta$, not like a gapless $p$- or $d$-wave $\Delta$.
Since the super-high-$T_c$ superconductors have shorter superconducting coherence length $\xi$ $(\xi \propto 1/\Delta \propto 1/T_c)$ of the Cooper pair with increasing the $\Delta$ and $T_c$, the superconducting fluctuations become stronger with increasing the $T_c$.
Therefore, at least $T_c >1,000$ K is necessary for the more stable and safe industrial applications of the superconductors at the room temperature (approximately more than three times higher $T_{\rm c}$ than the device operating room temperatures $\sim$300 K in order to suppress the superconducting fluctuations). 

One may have a serious question that it is possible to realize the macroscopic and coherent quantum effect in such a high temperature ($\geq$1,000 K), because the macroscopic quantum effect appears always at a low temperature such as the conventional low-$T_c$ superconductivity or Bose-Einstein condensation (BEC) in $^4$He.
One of the encouraging example in the real materials is the macroscopic and coherent quantum effect in the ferromagnetism of Fe and Co with super-high Curie temperature $(T_{\rm C})$, where $T_{\rm C} =1,043\ {\rm K}$ for Fe and $T_{\rm C} =1,388\ {\rm K}$ for Co.
The microscopic origin of the ferromagnetism is the repulsive intra-atomic Coulomb interaction ($U_{\rm eff}>0$ and order of a few eV) in the Jacques Friedel's localized virtual bound states (VBS) of the 3$d$ orbitals, and the kinetic energy gain by the hopping of the electrons between the VBS in the itinerant Fermion system with the partially occupied narrow 3$d$ band of the band width ($W$) to stabilize the ferromagnetism $(U_{\rm eff}>W)$ in the quantum theory of ferromagnets.
Then, the ferromagnetism shows the macroscopic and coherent quantum effect even at such a high temperature.
Based on the analogy of (1) the super-high-$T_{\rm C}$ in the ferromagnetic Fe and Co with macroscopic and coherent quantum effect caused by the repulsive intra-atomic Coulomb interaction $(U_{\rm eff}>0)$ in the VBS of the narrow 3$d$-band width $W$ $(U_{\rm eff} >W)$, and (2) the super-high-$T_c$ superconductors caused by the attractive Fermion system $(U_{\rm eff}<0)$ by a purely electronic mechanism, we can propose the knowledge-based and unified general design rules for a new CMD$^\regmark$ methodology to design and realize the super-high-$T_c$ superconductors.
Therefore, if it is possible to design and find the large ($\sim$eV) attractive electron-electron interactions in the itinerant Fermion systems based on the quantum theory, a possibility to realize a super-high-$T_c$ superconductor as a macroscopic and coherent quantum effect increases.

Our problem to be solved by the knowledge-based CMD$^\regmark$ is how we can search and design such a super-high-$T_c$ superconducting material with $s$-wave $\Delta$ from {\it ab initio} calculations using the atomic numbers only as an input parameter, without any experiments and empirical parameters.
For the realization of the super-high-$T_c$ superconductors, we need to design and find an attractive electron-electron interaction with the order of eV ($\approx$10,000 K).
This can be possible only based on purely electronic mechanisms and the searching for itinerant and attractive Fermion systems using {\it ab initio} calculations, since the quasi-particle mediated attractive-interaction mechanism such as the electron-phonon interaction by phonons (characteristic energy is around 50 meV) or spin-fluctuation mechanism mediated by magnons (characteristic energy is also around 50 meV) gives always too low-$T_c$ (maximum $T_c \approx 50\sim100\ {\rm K}$) compared with the room temperature or gapless $d$-wave superconductors, originating from the highly-correlated repulsive electron-electron interactions.
For the CMD$^\regmark$ of the super-high-$T_c$ superconductors, we should find and realize the very large attractive electron-electron interaction ($U_{\rm eff} < 0$) in the itinerant and attractive Fermion system upon the doping, so called negative effective $U$ $(U_{\rm eff}<0)$ system, where $U_{\rm eff} = E(N+1)+E(N-1)-2E(N) < 0$ and $E(N)$ is the total energy of a $N$ electron system, on the basis of {\it ab initio} calculation~\cite{HKY_16,HKY_17,Anderson_18,Fukushima_19}.
Then, $T_c$ and the phase diagram can be calculated by using the quantum Monte Carlo simulation or analytic theory based on the negative $U_{\rm eff}$ Hubbard model~\cite{Scalettar_20,Moreo_21,Nozieres_23} mapped from {\it ab initio} electronic structure calculations~\cite{Fukushima_19}.

Recently, Fukushima {\it et al.}~\cite{Fukushima_19} proposed the knowledge-based and unified general design rules for the negative $U_{\rm eff}$ systems by utilizing the purely electronic and attractive Fermion mechanisms:
(i) {\it the charge-excitation-induced negative $U_{\rm eff}$}~\cite{HKY_17,Fukushima_19}
and 
(ii) {\it the exchange-correlation-induced negative $U_{\rm eff}$}~\cite{HKY_16,Fukushima_19}. 
They applied the unified general design rules (above (i) and (ii) rules) to a realistic material by choosing the hole doped three-dimensional (3D) Chalcopyrite of CuFeS$_2$ using {\it ab initio} calculations, and reported the negative $U_{\rm eff}$ value of $-0.44$ eV.
A general phase diagram of the hole-doped CuFeS$_2$ was also predicted based on the results of the spin-polarized total energy calculations upon the hole doping.
Additionally, they proposed a generalized new searching methodology of the super-high-$T_c$ superconductor by the successive three steps (STEP 1, STEP 2, and STEP 3, which are called HOP, STEP and JUMP) using {\it ab initio} total energy and electron-phonon interaction calculations starting from the atomic numbers only as an input parameter, without any empirical parameters~\cite{Fukushima_19}. 

In this paper, we calculate the total energies of the different charge states of $E(N-1)$, $E(N)$ and $E(N+1)$ for the hole-doped 2D Delafossite of CuAlO$_2$, AgAlO$_2$ and AuAlO$_2$ by {\it ab initio} calculations, and find the large negative $U_{\rm eff}$ upon the hole doping in order to design the attractive and itinerant Fermion system for the realization of super-high-$T_c$ superconductors.
The itinerant and attractive Fermion system by negative $U_{\rm eff}$ is designed by
{\it the charge-excitation-induced negative $U_{\rm eff}$ mechanism}~\cite{HKY_17,Fukushima_19}, 
which is caused by the instability of the hole-doped chemical bonds of Cu$^{2+}(3d^9)$, Ag$^{2+}(4d^9)$, Au$^{2+}(5d^9)$ and O$^-(2p^5)$ relative to the un-doped closed shells of stable Cu$^+(3d^{10})$, Ag$^+(4d^{10})$, Au$^+(5d^{10})$ and O$^{2-}(2p^6)$.
Combining with the attractive electron-phonon interaction with $s$-wave superconductors $(T_c\sim50\ {\rm K})$ and the large attractive electron-electron interactions from negative $U_{\rm eff}$ ($U_{\rm eff} =-4.53$ eV for CuAlO$_2$, $U_{\rm eff} = -4.88$ eV for AgAlO$_2$ and $U_{\rm eff} = -4.14$ eV for AuAlO$_2$ by LDA),
we propose the general design rules for the realization of the super-high-$T_c$ superconductor for the future calculations using the multi-scale simulation combined with {\it ab initio} electronic structure calculations and quantum Monte Carlo simulations or analytic theory of the negative $U_{\rm eff}$ Hubbard model.
We also discuss the chemical trend and the dimensionality dependence of the negative $U_{\rm eff}$ system between the 2D vs. 3D structures by comparing the differences in the screening: the $p$-$d$ hybridization and coordination numbers between the 2D Delafossite of CuAlO$_2$, AgAlO$_2$ and AuAlO$_2$ and 3D Chalcopyrite of CuFeS$_2$.

This paper is organized as follows.
In Sec.\ \ref{sec:mo}, the relation between the missing oxidation states in the Periodic Table and the negative $U_{\rm eff}$ systems is discussed based on the physical and chemical points of view.
In Sec.\ \ref{sec:General_rule}, the unified general design rules for the purely electronic mechanisms of the attractive and itinerant Fermion system caused by negative $U_{\rm eff}$ are presented.
In Sec.\ \ref{sec:Strategies}, we propose the strategies how to design super-high-$T_c$ superconductors from {\it ab initio} calculations according to the unified general design rules~\cite{Fukushima_19}.
Section \ref{sec:Method} is devoted to the calculation details.
Our unified general design rules, which is based on the charge-excitation induced negative $U_{\rm eff}$ mechanism, are applied to the hole-doped 2D Delafossite of CuAlO$_2$, AgAlO$_2$ and AuAlO$_2$ in Sec.\ \ref{sec:Results}.
In Sec.\ \ref{sec:BCS_BEC}, by comparing the present results of the hole-doped of 2D Delafossite of the CuAlO$_2$, AgAlO$_2$ and AuAlO$_2$ and the previous results of the negative $U_{\rm eff}$ system in the hole-doped 3D Chalcopyrite CuFeS$_2$~\cite{Fukushima_19}, we discuss the chemical trends and dimension dependence of the evolution from weak- to strong-coupling superconductivity, which is continuous between Bardeen-Cooper-Schrieffer (BCS) and Bose-Einstein condensation (BEC) with increasing $|U_{\rm eff}|$ by changing the chemical compound, based on the theory of Nozi\'{e}res and Schmitt-Rink (hereafter, we call N-SR theory)~\cite{Nozieres_23}.
In Sec.\ \ref{sec:Summary}, finally, we summarize the present research and future prospects.

\section{\label{sec:mo}Negative $U_{\rm eff}$ System and Missing Oxidation States (Missing Oxidation Numbers) \protect\\}
In order to design and create the negative $U_{\rm eff}$ system upon the doping, we first explain the normal positive $U_{\rm eff}$ system comparing with the negative $U_{\rm eff}$ system.
The positive $U_{\rm eff}$ system means that the total energy $E(n)$ shows concavity downward as a function of the occupation numbers of electron ($n=N-1$, $N$ and $N+1$), and this downward concavity guarantees the stability of the $n=N$ electron system against the charge disproportionation or charge fluctuation of the two of $n=N$ electronic configurations into the $n=N-1$ and $n=N+1$ electronic configurations, where $2E(N) +U_{\rm eff} =E(N+1)+E(N-1)$ and $U_{\rm eff}>0$~\cite{Fukushima_19}.
The observed charge state of $N$ electron system in the experiments, such as Electron Paramagnetic Resonance (EPR) or Deep Level Transient Spectroscopy (DLTS) in the thermal equilibrium condition, show that the $N$ electron system has the repulsive electron-electron interaction ($U_{\rm eff}>0)$, where the charge state $(n=N)$ is stable forever against the charge fluctuation or charge disproportionation in the vicinity of $n=N$ electron system.

Due to the some special reasons originating from the purely electronic and attractive electron-electron interaction mechanisms~\cite{HKY_16,HKY_17,Fukushima_19}, such as
\begin{description}
\item[(1)] the charge-excitation-induced negative $U_{\rm eff}$ caused by the energy gain from the stability of the chemical bond in the closed shells~\cite{HKY_17,Fukushima_19},
\item[(2)] the exchange-correlation-induced negative $U_{\rm eff}$ caused by the energy gain from the Hund's rules in the high-spin ground states~\cite{HKY_16,Fukushima_19},
\vspace{0.5cm}

or negative $U_{\rm eff}$ system by the Jahn-Teller interaction caused by the strong electron-phonon ({\it local phonons}) interaction mechanism\cite{Anderson_18}

\vspace{0.5cm}
\item[(3)] the strong electron-phonon-interaction-induced negative $U_{\rm eff}$ caused by the energy gain from the Jahn-Teller effect with the local lattice-distortions (local phonons), which is called Anderson's negative $U_{\rm eff}$~\cite{Anderson_18}.
\end{description}
If the $n=N-1$ or $n=N+1$ electronic configuration becomes more stable than the $n=N$ electron configuration, one can realize that the total energy $E(n)$ shows convexity upward as a function of $n$, leading to the negative $U_{\rm eff}$ system in the thermal equilibrium condition.
In these cases, due to the stabilization of charge states in $n=N+1$ and $n=N-1$ electron configurations compared with the $2E(N)$ in $n=N$ electronic configuration, the charge disproportionation in the negative $U_{\rm eff}$ system can be expected: the two of the $N$ electron configurations makes charge disproportionation into the $n=N+1$ and $n=N-1$ electron configurations, where $2E(N){\rightarrow}E(N+1)+E(N-1)+|U_{\rm eff}|$ and $U_{\rm eff}< 0$.
In the negative $U_{\rm eff}$ system, the static charge disproportionation generally forms the charge density wave (CDW) in the insulating phases.
However, in the itinerant and attractive Fermion system upon the carrier doping, the BCS superconductivity or BEC caused by the attractive Fermion system can be stabilized~\cite{Fukushima_19,Nozieres_23}, depending on the relation of $|U_{\rm eff}|$ and $W$.
If $|U_{\rm eff} | > W$ in the negative $U_{\rm eff}$ system, with lowering temperature, the BEC is realized by forming the strong bound states of the Cooper pairs above $T_{\rm c}$ with the short coherence length $\xi$ ($\xi<l$, where $l$ is the carrier mean-free-path) and spin-singlet $(S=0)$ in the strong coupling limit according to the N-SR theory~\cite{Nozieres_23}.
On the other hand, if $|U_{\rm eff} |<W$ in the negative $U_{\rm eff}$ system, the BCS superconductivity is expected with spin-singlet $(S=0)$, where $T_c \propto \exp [-1/ |U_{\rm eff}| ]$ due to the weak coupling limit.

\begin{figure*}[t]
\begin{center}
\includegraphics[width=12cm,clip]{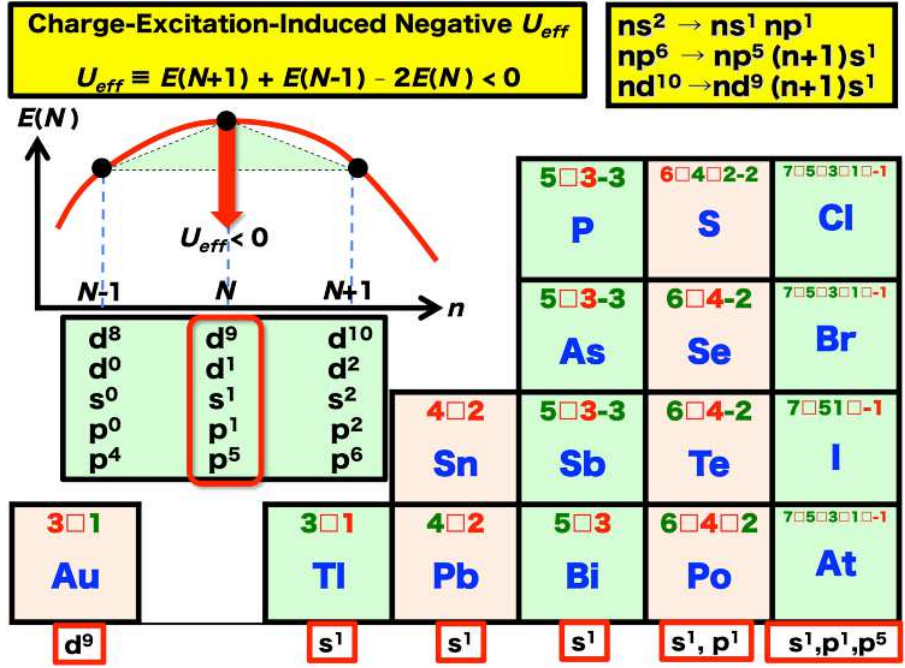}
\caption{Missing oxidation states ({\it missing oxidation numbers}) are shown by red square in the Periodic Tables~\cite{PeriodicTable_22}.
The numbers of existing oxidation states are shown in this figure.
The red-color number is the most important (often appeared in the nature) oxidation number in the compounds.
The missing oxidation states are in the $s^1$, $p^1$, $p^5$, $d^1$ and $d^9$ electronic configurations~\cite{HKY_17}, and these are located next to the stable closed shells of $s^2$, $p^6$, $d^0$ and $d^{10}$ electronic configurations.
The missing oxidation states mean that the oxidation states are unstable in the thermal equilibrium conditions, and indicate the negative $U_{\rm eff}$ systems with upward convexity of the total energy as a function of electron number $n$ as depicted in the upper-left figure.
If the $N$-electron system shows the negative $U_{\rm eff}$ ($U_{\rm eff}<0$), the $N$-electron system becomes unstable with upward convexity in the total energy as a function of electron occupations $n$, then disproportionate into the $(N+1)$- and $(N-1)$-electron systems $(2E(N){\rightarrow}E(N+1)+E(N-1)+|U_{\rm eff}|)$ by gaining the energy of $|U_{\rm eff}|$, as was shown schematically in the upper-left figure.
Upper-right table shows the transition from the ground states of the closed shells to the first excited states, which stabilize the closed shells through the second order perturbation.}
\label{mo}
\end{center}
\end{figure*}

In order to design super-high-$T_c$ superconductors, we proposed the unified general design rules for searching and designing the negative $U_{\rm eff}$ system by (i) {\it the charge-excitation-induced negative $U_{\rm eff}$ mechanism}~\cite{HKY_17,Fukushima_19}, which is strongly correlated to the experimentally observed missing oxidation states (sometimes {\it missing oxidation numbers}) as the digital-data bases in the Periodic Table~\cite{PeriodicTable_22}.
The Periodic Table (see Fig.\ \ref{mo}) shows that the missing oxidation states exist in the some periodicity and rules with the simple regulations related to the chemical bond stability in the closed shells.
For example, the following oxidation states are missing in the Periodic Table, even if the both sides of the adjacent oxidation states exist: Au$^{2+}(d^9)$, Nb$^{4+}(d^1)$, V$^{4+}(d^1)$, Ti$^{3+}(d^1)$, Tl$^{2+}(s1)$, Sn$^{3+}(s^1)$, Pb$^{3+}(s^1)$, P$^{4+}(s^1)$, As$^{4+}(s^1)$, Sb$^{4+}(s^1)$, Bi$^{4+}(s^1)$, S$^{5+}(s^1)$, S$^{3+}(s^2p^1)$, Se$^{5+}(s^1)$, Te$^{5+}(s^1)$, Po$^{5+}(s^1)$, Po$^{3+}(s^2p^1)$, Cl$^{6+}(s^1)$, Cl$^{4+}(s^2p^1)$, Cl$^{2+}(p^5)$, Cl$^0(s^2p^5)$, Br$^{6+}(s^1)$, Br$^{4+}(s^2p^1)$, Br$^{2+}(p^5)$, Br$^0(s^2p^5)$, I$^{6+}(s^1)$, I$^0(s^2p^5)$, At$^{6+}(s^1)$, At$^{4+}(s^2p^1)$, At$^{2+}(p^5)$ and At$^0(s^2p^5)$ (see Fig.\ \ref{mo}, Refs.\ \cite{HKY_17}, \cite{Fukushima_19} and \cite{PeriodicTable_22}).

The missing oxidation states correspond to the negative $U_{\rm eff}$ systems, since the missing oxidation states with two of $N$ electron configurations make charge disproportionation into $N+1$ and $N-1$ electron configurations in the thermal equilibrium condition: $2E(N){\rightarrow}E(N+1)+E(N-1)+|U_{\rm eff}|$ by gaining the energy of $|U_{\rm eff}|$ in the case of the negative $U_{\rm eff}$ system.
Therefore, one can find the following charge disproportionation for the ions with the missing oxidation states in the Periodic Table:

\begin{eqnarray}
2{\rm Au}^{2+}(d^9) &\rightarrow& {\rm Au}^+(d^{10})+{\rm Au}^{3+}(d^8)+|U_{\rm eff}|,\nonumber\\
2{\rm Nb}^{4+}(d^1)  &\rightarrow&     {\rm Nb}^{3+}(d^2)   +  {\rm Nb}^{5+}(d^0)   +  |U_{\rm eff}|,\nonumber\\
    2{\rm V}^{4+}(d^1)   &\rightarrow&     {\rm V}^{3+}(d^2)    +   {\rm V}^{5+}(d^0)   +  |U_{\rm eff}|,\nonumber\\
    2{\rm Ti}^{3+}(d^1)   &\rightarrow&    {\rm Ti}^{2+}(d^2)    +   {\rm Ti}^{4+}(d^0)   +  |U_{\rm eff}|,\nonumber\\
2{\rm Tl}^{2+}(s^1)    &\rightarrow&   {\rm Tl}^+(s^2)    +  {\rm Tl}^{3+}(s^0)   +  |U_{\rm eff}|,\nonumber\\
2{\rm Sn}^{3+}(s^1)    &\rightarrow&   {\rm Sn}^{2+}(s^2)   +  {\rm Sn}^{4+}(s^0)  +   |U_{\rm eff}|,\nonumber\\ 
     2{\rm Pb}^{3+}(s^1)    &\rightarrow&   {\rm Pb}^{2+}(s^2)   +  {\rm Pb}^{4+}(s^0)  +   |U_{\rm eff}|,\nonumber\\
     2{\rm As}^{4+}(s^1)    &\rightarrow&   {\rm As}^{3+}(s^2)   +  {\rm As}^{5+}(s^0)  +   |U_{\rm eff}|,\nonumber\\
     2{\rm Sb}^{4+}(s^1)    &\rightarrow&   {\rm Sb}^{3+}(s^2)   +  {\rm Sb}^{5+}(s^0)  +  |U_{\rm eff}|,\nonumber\\
     2{\rm Bi}^{4+}(s^1)    &\rightarrow&   {\rm Bi}^{3+}(s^2)    +  {\rm Bi}^{5+}(s^0)  +   |U_{\rm eff}|,\nonumber\\
     2{\rm P}^{4+}(s^1)    &\rightarrow&   {\rm P}^{3+}(s^2)    +  {\rm P}^{5+}(s^0)  +   |U_{\rm eff}|,\nonumber\\
     2{\rm S}^{5+}(s^1)     &\rightarrow&   {\rm S}^{4+}(s^2)     +  {\rm S}^{6+}(s^0)  +   |U_{\rm eff}|,\nonumber\\
     2{\rm S}^{3+}(s^2p^1)   &\rightarrow&   {\rm S}^{2+}(s^2p^2)    +  {\rm S}^{4+}(s^2p^0)  +  |U_{\rm eff}|,\nonumber\\
     2{\rm Se}^{5+}(s^1)    &\rightarrow&   {\rm Se}^{4+}(s^2)    +  {\rm Se}^{6+}(s^0)  +   |U_{\rm eff}|,\nonumber\\
     2{\rm Te}^{5+}(s^1)    &\rightarrow&   {\rm Te}^{4+}(s^2)    +  {\rm Te}^{6+}(s^0)  +   |U_{\rm eff}|,\nonumber\\
     2{\rm Po}^{5+}(s^1)    &\rightarrow&   {\rm Po}^{4+}(s^2)    +  {\rm Po}^{6+}(s^0)  +   |U_{\rm eff}|,\nonumber\\
     2{\rm Po}^{3+}(s^2p^1)  &\rightarrow&   {\rm Po}^{2+}(s^2p^2)   +  {\rm Po}^{4+}(s^2p^0) +   |U_{\rm eff}|,\nonumber\\
     2{\rm Cl}^{6+}(s^1)    &\rightarrow&   {\rm Cl}^{5+}(s^2)    +   {\rm Cl}^{7+}(s^0)  +   |U_{\rm eff}|,\nonumber\\
     2{\rm Cl}^{4+}(s^2p^1)  &\rightarrow&   {\rm Cl}^{3+}(s^2p^2)   +  {\rm Cl}^{5+}(s^2p^0)  + |U_{\rm eff}|,\nonumber\\
     2{\rm Cl}^{2+}(s^0p^5)  &\rightarrow&   {\rm Cl}^+(s^0p^6)    +  {\rm Cl}^{3+}(s^0p^4)  +  |U_{\rm eff}|,\nonumber\\
     2{\rm Cl}^0(s^2p^5)   &\rightarrow&   {\rm Cl}^-(s^2p^6)    +  {\rm Cl}^{+}(s^2p^4)  +  |U_{\rm eff}|,\nonumber\\
     2{\rm Br}^{6+}(s^1)    &\rightarrow&   {\rm Br}^{5+}(s^2)    +  {\rm Br}^{7+}(s^0)   +   |U_{\rm eff}|,\nonumber\\
     2{\rm Br}^{4+}(s^2p^1)  &\rightarrow&   {\rm Br}^{3+}(s^2p^2)   +  {\rm Br}^{5+}(s^2p^0)  +  |U_{\rm eff}|,\nonumber\\
     2{\rm Br}^{2+}(s^0p^5)  &\rightarrow&   {\rm Br}^+(s^0p^6)    +  {\rm Br}^{3+}(s^0p^4)  + |U_{\rm eff}|,\nonumber\\
     2{\rm Br}^0(s^2p^5)   &\rightarrow&   {\rm Br}^-(s^2p^6)    +   {\rm Br}^+(s^2p^4)  +  |U_{\rm eff}|,\nonumber\\
     2{\rm I}^{6+}(s^1)     &\rightarrow&   {\rm I}^{5+}(s^2)      +   {\rm I}^{7+}(s^0)    + |U_{\rm eff}|,\nonumber\\
     2{\rm I}^0(s^2p^5)    &\rightarrow&   {\rm I}^-(s^2p^6)      +  {\rm I}^+(s^2p^4)    + |U_{\rm eff}|,\nonumber\\
     2{\rm At}^{6+}(s^1)   &\rightarrow&   {\rm At}^{5+}(s^2)      +  {\rm At}^{7+}(s^0)    +  |U_{\rm eff}|,\nonumber\\
     2{\rm At}^{4+}(s^2p^1)  &\rightarrow&  {\rm At}^{3+}(s^2)      +   {\rm At}^{5+}(s^0)   +  |U_{\rm eff}|,\nonumber\\
     2{\rm At}^{2+}(s^0p^5)  &\rightarrow&  {\rm At}^+(s^0p^6)     +   {\rm At}^{3+}(s^0p^4)  +  |U_{\rm eff}|,\nonumber\\
     2{\rm At}^0(s^2p^5)  &\rightarrow&  {\rm At}^-(s^2p^6)     +   {\rm At}^+(s^2p^4)  +   |U_{\rm eff}|.\nonumber
\end{eqnarray}

In the negative $U_{\rm eff}$ system with the insulating compounds or elements, the static charge disproportionation can be observed according to the above reactions.
However, if the itinerant and attractive Fermion system with the large negative $U_{\rm eff}$ upon the hole doping (or electron doping) is stabilized by utilizing the more fluctuation-induced system by the low-dimensional system to avoid the static charge disproportionation such as the CDW, one can realize the BCS superconductivity or BEC with a large dynamical charge-fluctuation based on the attractive electron-electron interactions in the itinerant Fermion system.

\section{\label{sec:General_rule}Unified General Design Rules of the Negative $U_{\rm eff}$ for the Realization of Itinerant and Attractive Fermion System\protect\\}
Here, we propose the unified general design rules for the missing oxidation states based on the stability of the chemical bond in the closed shells and the instability of the hole-doped (or electron-doped) chemical bond.
The existing oxidation states and the missing oxidation states in the Periodic Table is one of the most important digital-data bases which reflect the nature of the physical stability and chemical properties in the natural or artificially made chemical compounds.
The electronic configurations of the missing oxidation states do not exist in the chemical compounds and elements in the thermal equilibrium conditions.
Therefore, the element with the negative $U_{\rm eff}$ corresponds to the missing oxidation states in the Fig.\ \ref{mo}, and the compound with the negative $U_{\rm eff}$ system becomes unstable and forming a charge disproportionation in the thermal equilibrium conditions:
\begin{description}
\item[(1)] the strong static charge disproportionation caused by the CDW formation in the insulating phases or
\item[(2)] the superconductivity caused by the dynamical charge-fluctuation by the attractive electron-electron interactions forming Cooper pairs with spin-singlet $(S=0)$, leading to the BCS superconductivity in the weak coupling regime $( |U_{\rm eff} |<W )$, or BEC caused by the composite Bosons by the strong bound states with short $\xi$ and spin-singlet $(S=0)$ in the strong coupling regime $( |U_{\rm eff} |>W )$. 
\end{description}
All of the missing oxidation states correspond to the $s^1$, $s^2p^1$, $s^2p^5$, $s^0p^5$, $d^1$ and $d^9$ electronic configurations on the Periodic Tables~\cite{PeriodicTable_22}, which are located next to (or just before) the closed shells of stable electronic configurations, such as the $s^2$, $p^6$, $d^0$ and $d^{10}$ electronic configurations.

The stability of the closed shells is caused by the strong hybridization between the ground states of the closed shells (such as the $s^2$, $p^6$, $d^0$ and $d^{10}$ electronic configurations) and the first excited states (such as the $s^1p^1$, $p^5s^1$, $p^5d^1$ and $d^9s^1$ electronic configurations).
The energy of the first excited state in the closed shell is very close to that of the ground state (such as $ns^2 \rightarrow ns^1np^1$, $nd^{10} \rightarrow nd^9(n+1)s^1$, $np^6 \rightarrow np^5(n+1)s^1$, where $n$ is the principal quantum number, {\it e.g.}, the ground states of Cu$^+(3d^{10})$, Ag$^+(4d^{10})$, Au$^+(5d^{10})$, O$^{2-}(2p^6)$ and first excited states of Cu$^+(3d^94s^1)$, Ag$^+(4d^95s^1)$, Au$^+(5d^96s^1)$, O$^{2-}(2p^53s^1)$, respectively).
Then, these two states repel very strongly through the second order perturbation by the charge-excitation-induced mechanism, so that the ground states of the closed shells is stabilized dramatically
(this is the physical and quantum chemical mechanism to explain the empirical chemical rule of the chemical-bond stability in the closed shells).
Therefore, the negative $U_{\rm eff}$ system just before (or just after) the electronic configurations of the closed shells can be realized by the hole doping or electron doping.
Based on the above charge-excitation-induced negative $U_{\rm eff}$ mechanism, one can obtain the upward convexity in the total energy for $n=N$ electron configurations, which is located next to the stable closed shell of $n=N+1$ or $n=N-1$ electron configurations, {\it i.e.}, the charge-excitation-induced negative $U_{\rm eff}$ system for $n=N$ electron configurations (such as the $s^1$, $s^2p^1$, $s^2p^5$, $d^1$ and $d^9$ electronic configurations, see Fig.~\ref{mo}).

Two types of the unified general design rules by purely electronic mechanism of the negative $U_{\rm eff}$ are already proposed~\cite{HKY_16,HKY_17,Fukushima_19} based on, 
\begin{description}
\item[(1)] the charge-excitation-induced negative $U_{\rm eff}$ for $s^1$, $p^1$, $p^5$, $d^1$ and $d^9$ electronic configurations~\cite{HKY_17,Fukushima_19} caused by the stability of chemical bond in the closed shells of $s^2$, $p^6$, $d^0$ and $d^{10}$ electronic configurations,
\item[(2)] the exchange-correlation-induced negative $U_{\rm eff}$ for $d^4$ and $d^6$ electronic configurations~\cite{HKY_16,Fukushima_19} due to the stability of $d^5$ electronic configuration in the Hund's rules with the high-spin states.
\end{description}

In 2008, Katayama-Yoshida {\it et al.}~\cite{HKY_17} proposed the charge-excitation-induced negative $U_{\rm eff}$ system, where the additional total energy of $E(N-1)$ electron system in $s^0$ (or $p^0$, $p^4$, $d^0$, $d^8$) electronic configuration and $E(N+1)$ electron system in $s^2$ (or $p^2$, $p^6$, $d^2$, $d^{10}$) electronic configuration becomes more stable than the two of $N$ electron system in $s^1$ (or $ p^1$, $p^5$, $d^1$, $d^9$) electronic configurations by lowering $E(N)$ through mixing of ground state $s^2$ (or $p^2$, $p^6$, $d^2$, $d^{10}$) electronic configuration with the first excited states.
In the reaction of the charge-excitation-induced negative $U_{\rm eff}$ system, $2s^1 \rightarrow s^0 + s^2+|U_{\rm eff}|$  (or $2p^1 \rightarrow p^0 + p^2+|U_{\rm eff}|$, $2p^5 \rightarrow p^4 + p^6+|U_{\rm eff}|$, $2d^1 \rightarrow d^0 + d^2+|U_{\rm eff}|$, $2d^9 \rightarrow d^8 + d^{10}+|U_{\rm eff}|$), we can expect a charge disproportionation and insulating CDW in the stoichiometric compounds. 
However, if we can realize the charge-excitation-induced negative $U_{\rm eff}$ system with itinerant Fermion gas upon $p$- or $n$-type doping or under the applied ultra-high pressures by destabilizing CDW through the pressure-induced chemical reaction, we can stabilize a superconducting state by Cooper pairs or BEC by the composite Bosons through attractive carrier dynamics of the Fermion system.
In experiment, the existence of the missing oxidation states of ions, such as Tl$^{2+}(6s^1)$, Sn$^{3+}(5s^1)$, Pb$^{3+}(6s^1)$, P$^{4+}(3s^1)$, As$^{4+}(4s^1)$, Sb$^{4+}(5s^1)$, Hg$^+(6s^1)$, Bi$^{4+}(6s^1)$, S$^{3+}(3s^23p^1)$, S$^{5+}(3s^1)$, Se$^{5+}(4s^1)$, Te$^{5+}(5s^1)$ and Po$^{5+}(6s^1)$, in the Periodic Table is well known.
These atoms indicate the charge-excitation-induced negative $U_{\rm eff}$ system, where the charge disproportionation occurs in the insulating compounds.
Therefore, all of these insulating compounds such as Ba$^{2+}(s^0)$Bi$^{4+}(s^1)$O$^{2-}_3(s^2p^6)$ and Bi$^{4+}(s^1)$Fe$^{2+}(d^6)$O$^{2-}_3(s^2p^6)$ indicate the insulating CDW (ferroelectricity) caused by the charge-excitation-induced negative $U_{\rm eff}$.

In 1985, Katayama-Yoshida and Zunger~\cite{HKY_16} proposed the exchange-correlation-induced negative $U_{\rm eff}$ system for $d^4$ and $d^6$ electronic configurations based on the stability of $d^5$ electronic structures by the Hund's rules with high-spin states for the magnetic 3$d$-transition metal atom impurities in semiconductors.
The key point of the exchange-correlation-induced negative $U_{\rm eff}$ system is that when a free 3$d$-transition atom is placed into the polarizable materials, its Coulomb (charge) and exchange-correlation interaction (spin) respond in fundamentally different ways to the screening.
The Coulomb interaction in the charge is responding to long-wavelength ({\it monopole screening}), and is reduced far more than the exchange-correlation ({\it multi-pole screening}).
Therefore, one can see that the local charge of 3$d$-transition atom changes very small amount upon the doping and realized the different charge states, however, the local magnetic moment of 3$d$-transition atom caused by the exchange-correlation interaction changes dramatically by changing the charge states ({\it oxidation states measured by DLTS or EPR experiment}) upon the doping~\cite{HKY_16}.
Here, we can say that the charge and spin degree of the freedom is screened completely different way in the polarizable materials upon the doping.
These are very important functionality to control the charge and spin degrees of freedom in the condensed matter upon the doping or electric field in order to design the new high-$T_c$ superconductors by creating the itinerant and attractive Fermion system.
The exchange-correlation-induced negative $U_{\rm eff}$ for the realization of attractive Fermion system is purely electronic mechanism, therefore, we can design a large negative effective $U_{\rm eff}$ system ($\sim$eV) based on the exchange-correlation interactions~\cite{Fukushima_19}.

In 1975, Anderson~\cite{Anderson_18} proposed negative $U_{\rm eff}$ system in the Chalcogenide-glass or amorphous semiconductors to explain the missing oxidation states in the EPR experiments.
The Anderson's negative $U_{\rm eff}$ system is caused by the energy gain by Jahn-Teller interaction and is not based on the purely electronic mechanism but caused by the strong electron-phonon interactions ({\it local phonons}).
Many defects and impurities showing Anderson's negative $U_{\rm eff}$ system are discovered in 1980s, and it became very popular for the various defects or impurities in semiconductors and ionic materials, such as silicon-vacancy and boron interstitials in silicon (see the References in the Ref.\ \cite{HKY_16}).
Since the Anderson's negative $U_{\rm eff}$ system originates from the Jahn-Teller energy gain from the local phonons of defects by Jahn-Teller interactions, we can observe the metastable negative $U_{\rm eff}$ defects or negative $U_{\rm eff}$ impurities by the EPR experiment at the very low temperature ({\it by freezing the local-phonons at the low temperature}) by creating the metastable negative $U_{\rm eff}$ defects ({\it missing oxidation states}) through the photo-excitations of electrons.
In order to realize the super-high-$T_c$ superconductors with $T_c>1,000$ K, we do need to design a large negative $U_{\rm eff}$ system which can be only realized by the purely electronic and attractive Fermion mechanisms with the order of $\sim$eV~\cite{HKY_16,HKY_17,Fukushima_19}.

As was shown in previous design papers of negative $U_{\rm eff}$ system~\cite{HKY_17,Fukushima_19}, we always expect the competition between the CDW phase and BCS superconducting phase (or BEC phase depending on the relation between the $|U_{\rm eff}|$ and $W$) in the large negative $U_{\rm eff}$ system.
In order to stabilize the BCS superconducting phase or BEC phase relative to the insulating CDW phase, we need to create the itinerant and the attractive Fermion system upon the doping and to enhance the charge-fluctuations by controlling the low-dimensionality with the large negative $U_{\rm eff}$ system in the 2D Delafossite.
In the BCS superconducting phase or BEC phase in the 2D Delafossite, we also expect the large superconducting fluctuations due to the short coherence length of the superconducting states in the 2D Delafossite.
In order to realize the super-high-$T_c$ superconductors, we need to design and realize the large negative $U_{\rm eff}$ ($\sim$ eV $\approx$ $\sim$ 10,000 K) system to avoid the strong superconducting fluctuations at the room temperature in the 2D Delafossite.

In order to detect the negative $U_{\rm eff}$ system for the experimental confirmation, which is originated by purely electronic mechanisms (the charge-excitation-induced or the exchange-correlation-induced negative $U_{\rm eff}$ mechanisms), we can estimate the negative $U_{\rm eff}$ $(U_{\rm eff} = E(N+1) + E(N-1) - 2E(N) = \Delta E([N]/[N+1]) - \Delta E([N-1]/[N]) < 0)$ by measuring that the first ionization energy $(\Delta E([N-1]/[N]) = E(N)- E(N-1))$ is larger than that of the second ionization energy $(\Delta E([N]/[N+1]) = E(N+1)- E(N))$, then leading to the negative $U_{\rm eff}$.

\section{\label{sec:Strategies}New Strategies for the Design of Super-High-$T_c$ Superconductors based on the First Principles Calculations\protect\\}
In order to search and design for the super-high-$T_c$ superconductors with $s$-wave superconducting gap $(\Delta)$, we have proposed the following new universal design strategies for the realistic CMD$^\regmark$ and realization, which is consist of Three STEPS (by {\it HOP}, {\it STEP} and {\it JUMP})~\cite{Fukushima_19};
\begin{description}
\item[STEP 1 (HOP)] First, we should find the $s$-wave superconductors with $T_c >50 \sim 100\ {\rm K}$ caused by strong electron-phonon interactions.
In order to avoid the successive phase transition of the structures and lattices, we need to perform a new doping and valence control methods, which should stabilize the crystal structures and lattices upon the doping, such as the doping method where the holes go to the occupied anti-bonding states or doping method where the electrons go to the un-occupied bonding states.
Hole doping to the closed shells of occupied $d^{10}$ electronic configuration ({\it occupied anti-bonding states}) or electron doping to the unoccupied $d^0$ electronic configuration ({\it unoccupied bonding states}) is a good doping candidate for this purpose in order to stabilize the crystal structures and lattices upon the doping.
In this case, the electron-phonon interactions mediated $T_c$ can be correctly estimated or predicted by the conventional {\it ab initio} electronic structure calculations~\cite{Nakanishi_14,Nakanishi_15}.
In order to realize such a special compounds or elements upon the doping, we can not use the conventional semiconductor or ionic materials such as Si, Diamond, GaAs, ZnSe, ZnO, NaCl, CaF$_2$, {\it etc}, because in these conventional materials the valence band is bonding states ({\it then, doped holes go to the bonding state and weaken the covalency}) and the conduction band is anti-bonding states ({\it then, doped electrons go to the anti-bonding state and weaken the covalency}).
We should design the special compounds or elements, such as the Chalcopyrite or Delafossite structures with $d^{0}$ or $d^{10}$ electronic configurations, which satisfies the conditions to enhance the covalency upon the doping.
\item[STEP 2 (STEP)] We need to design of the negative $U_{\rm eff}$ system originating from the purely electronic mechanisms with the attractive Fermion interactions by negative $U_{\rm eff}$ with the order of $\sim$eV in addition to the {\it STEP 1}.
Possible candidates are
\begin{description}
\item[(1)] The charge-excitation-induced negative $U_{\rm eff}$ systems for $s^1$, $p^1$, $p^5$, $d^1$, $d^9$, $f^1$ and $f^{13}$ electronic configurations~\cite{HKY_17,Fukushima_19}, or/and
\item[(2)] The exchange-correlation-induced negative $U_{\rm eff}$ systems~\cite{HKY_16,Fukushima_19} for $d^4$, $d^6$, $f^6$, and $f^8$ electronic configurations based on the stability of high-spin state by the Hund's rules, which show the attractive electron-electron interactions in the order of $\sim$eV~\cite{Fukushima_19}. 
\end{description}
One can find these negative $U_{\rm eff}$ systems based on the total energy by {\it ab initio} electronic structure calculations for the different charge states, then the following charge disproportionation or dynamical charge fluctuation can be expected with the energy gain by $|U_{\rm eff}|$;
\begin{description}
\item[(1)] In the case of charge-excitation-induced negative $U_{\rm eff}$ systems,
\begin{eqnarray}
\hspace{-3cm}U_{\rm eff} = E(s^2) + E(s^0) - 2E(s^1) < 0, \hspace{0.2cm} &\rm then&, \hspace{0.2cm} 2E(s^1) \rightarrow  E(s^2) + E(s^0)  +  |U_{\rm eff}|,\nonumber\\
\hspace{-3cm}U_{\rm eff} = E(p^2) + E(p^0) - 2E(p^1) < 0, \hspace{0.2cm} &\rm then&, \hspace{0.2cm} 2E(p^1) \rightarrow  E(p^2)+E(p^0)  +|U_{\rm eff}|,\nonumber\\
\hspace{-3cm}U_{\rm eff} = E(p^6) + E(p^4) - 2E(p^5) < 0, \hspace{0.2cm} &\rm then&, \hspace{0.2cm} 2E(p^5) \rightarrow  E(p^6) + E(p^4)  +  |U_{\rm eff}|,\nonumber\\
\hspace{-3cm}U_{\rm eff} = E(d^2) + E(d^0) - 2E(d^1) < 0, \hspace{0.2cm} &\rm then&, \hspace{0.2cm} 2E(d^1) \rightarrow  E(d^2) + E(d^0)  +  |U_{\rm eff}|,\nonumber\\
\hspace{-3cm}U_{\rm eff} = E(d^{10}) + E(d^8) - 2E(d^9) < 0, \hspace{0.2cm} &\rm then&, \hspace{0.2cm} 2E(d^9) \rightarrow  E(d^{10}) + E(d^8)  +  |U_{\rm eff}|,\nonumber\\
\hspace{-3cm}U_{\rm eff} = E(f^{14}) + E(f^{12}) - 2E(f^{13}) < 0, \hspace{0.2cm} &\rm then&, \hspace{0.2cm} 2E(f^{13}) \rightarrow  E(f^{14}) + E(f^{12})  +  |U_{\rm eff}|,\nonumber\\
{\rm or/and}\nonumber
\end{eqnarray}
\item[(2)] In the case of the exchange-correlation-induced negative $U_{\rm eff}$ systems,
\begin{eqnarray}
\hspace{-3cm}U_{\rm eff} = E(d^5) + E(d^3)-2E(d^4) < 0,  \hspace{0.2cm} &\rm then&, \hspace{0.2cm}  2E(d^4) \rightarrow  E(d^5) + E(d^3)  +  |U_{\rm eff}|,\nonumber\\
\hspace{-3cm}U_{\rm eff} = E(d^7)+E(d^5)-2E(d^6) < 0,   \hspace{0.2cm} &\rm then&, \hspace{0.2cm}  2E(d^6) \rightarrow  E(d^5) + E(d^7)  +  |U_{\rm eff}|,\nonumber\\
\hspace{-3cm}U_{\rm eff} = E(f^7) + E(f^5) - 2E(f^6) < 0,  \hspace{0.2cm} &\rm then&, \hspace{0.2cm}  2E(f^6) \rightarrow  E(f^5) + E(f^7)   +  |U_{\rm eff}|,\nonumber\\
\hspace{-3cm}U_{\rm eff} = E(f^9) + E(f^7) - 2E(f^8) < 0,  \hspace{0.2cm} &\rm then&, \hspace{0.2cm}  2E(f^8) \rightarrow  E(f^9) + E(f^7)   +  |U_{\rm eff}|,\nonumber
\end{eqnarray}
\end{description}
where $E(s^n)$, $E(d^n)$ and $E(f^n)$ are the total energies of $s^n$, $d^n$ and $f^n$ electronic configurations with $n$ electron systems in the compounds, respectively. 
\item[STEP 3 (JUMP)] We can estimate the $T_c$ and calculate the phase diagram of the negative $U_{\rm eff}$ system by the quantum Monte Carlo simulation (without sign-problems because of the negative $U_{\rm eff}$ system) or the analytic theory such as the N-SR theory based on the negative $U_{\rm eff}$ Hubbard model~\cite{Nozieres_23}, which is mapped from {\it ab initio} calculations of the negative $U_{\rm eff}$, hopping integral $t$ and chemical potential $\mu$ upon the doping~\cite{Scalettar_20,Moreo_21}.
Based on {\it ab initio} electronic structure calculations, we should define the most important single band ({\it e.g.}, see the highest valence band near VBM in Fig.\ \ref{band}) and we can do fitting by the simple tight-binding band, and also calculate the $U_{\rm eff}$ from {\it ab initio} total energy calculations by changing the charge state upon the doping.
If we need to extend the model to the more realistic materials by taking into account the more bands in realistic way, we can easily extend to the multi-band negative $U_{\rm eff}$ Hubbard model.
Once, we obtained the negative $U_{\rm eff}$ Hubbard model from the first principles calculations, the relation between the BEC and BCS superconductivity can be predicted by the N-SR theory~\cite{Nozieres_23} correctly.
The evolution from weak- to strong-coupling superconductivity is smooth and seamless, and then the only crossover can be seen between BCS and BEC with increasing $|U_{\rm eff}|$, based on the N-SR theory.
In the theoretically proposed negative $U_{\rm eff}$ Hubbard model from the first principles, the relation between the $|U_{\rm eff}|$ and $W$ of negative $U_{\rm eff}$ Hubbard model is essentially important to discriminate the following BCS and BEC regimes; (1) {\it the BCS weak coupling regime} $(|U_{\rm eff}| < W )$, (2) {\it the BCS/BEC crossover regime} $(|U_{\rm eff}| \approx W )$ and
(3) {\it the BEC strong coupling regime} $(|U_{\rm eff}| > W )$.
We can predict that the $T_c$ increases exponentially $(T_c \propto \exp [-1/|U_{\rm eff}| ] )$ in (1), the $T_c$ goes through a maximum when $|U_{\rm eff}| \approx W$ in (2) and the $T_c$ decreases with increasing $|U_{\rm eff}|$ in (3), depending on the calculated value of $|U_{\rm eff}|$ and $W$ in the negative $U_{\rm eff}$ Hubbard model.
\end{description}

\section{\label{sec:Method}Calculation Method\protect\\}
Generally, we can calculate the total energy and band structure based on local density approximation (LDA) calculation or more advanced beyond LDA calculation by such as SIC-LDA~\cite{Toyoda_SIC} (self-interaction-corrected LDA), QSGW~\cite{Kotani_QSGW} (quasi-particle self-consistent GW method) or another more advance methodology based on the reliable conventional methods.
In this paper, the calculations were performed within the density functional theory (DFT)~\cite{Hohenberg_24,Kohn_25} with a plane-wave pseudopotential method, as implemented in the Quantum-ESPRESSO code~\cite{Giannozzi_26}.
We employed the Perdew-Wang 91 generalized gradient approximation (GGA) exchange-correlation functional~\cite{Perdew_27} and ultra-soft pseudopotentials~\cite{Vanderbilt_28}.
For the pseudopotentials, the 3$d$ electrons of the transition metals were also included in the valence electrons.
In reciprocal lattice space integral calculation, we used $8\times8\times8$ $k$-point grids in the Monkhorst-Pack grid~\cite{Monkhorst_29}.
The energy cut-off for wave function was 40 Ry and that for charge density was 320 Ry.
These $k$-point grids and cut-off energies are fine enough to achieve convergence within 10 mRy/atom in the total energy.
CuAlO$_2$, AgAlO$_2$ and AuAlO$_2$ are Delafossite structure, which belongs to the space group $R\bar{3}m$ (No.166) and is represented by the cell parameters $a$ and $c$ and the internal parameter $z$.
These parameters were optimized by the constant-pressure variable-cell relaxation using the Parrinello-Rahman method~\cite{Parrinello_30}.
After the relaxation, we calculated the total energy of the different charged states, and estimated the  $U_{\rm eff}$ value.
These structure optimization and total energy calculations were performed with total charge of the system (tot\_charge: input parameter of Quantum-ESPRESSO) = 0.0, 0.1, 0.2, ${\cdots}$, 2.0.
The total energy calculations were performed for the different charged states of $E(N-1)$, $E(N)$ and $E(N+1)$ electron systems for the hole-doped CuAlO$_2$, AgAlO$_2$ and AuAlO$_2$ by decreasing the number of electrons in the unit cell with keeping the charge neutrality by the jellium background charge in order to satisfy the real experimental condition upon the chemical doping from the charge reservoir.
For the hole-doped CuFeS$_2$ system, the ``Vienna Ab initio Simulation Package'' (VASP), based on 
projector augmented wave (PAW) pseudopotentials~\cite{Kresse}, 
was used.
Readers can see the calculation details in Ref.~\cite{Fukushima_19}.
It is well known that the main reason of the error of LDA~\cite{Toyoda_SIC}, such as band-gap energy, is caused by the self-interaction of electrons through the LDA.
In order to predict the correct band-gap and quasi-particle excitations, we implemented the SIC into first-principles calculation code (PSIC-LDA)~\cite{Nakanishi_12} and applied PSIC-LDA to the Delafossite of CuAlO$_2$, AgAlO$_2$ and AuAlO$_2$.
Our simulation shows that the valence band width calculated within the PSIC-LDA is narrower than that calculated within the only LDA.
The reason is that the PSIC-LDA makes the $d$-band potential more attractive.
The energy gap calculated within the PSIC-LDA expand and is close to experimental data~\cite{Nakanishi_12}.

\section{\label{sec:Results}Calculated Results and Discussion\protect\\}
\subsection{Calculated Electronic Structures of 2D Delafossite of CuAlO$_2$, AgAlO$_2$ and AuAlO$_{2}$}
\begin{figure}[t]
\begin{center}
\includegraphics[height=8cm,clip]{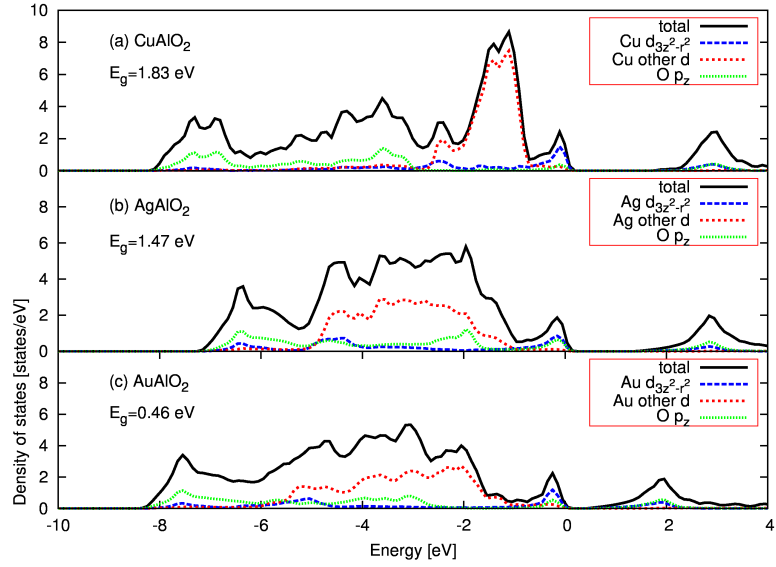}
\caption{Calculated total density of states (DOS) and partial DOS of the un-doped 2D Delafossite structures of (a) CuAlO$_2$, (b) AgAlO$_2$ and (c) AuAlO$_2$ by LDA.
The energies are measured from the valence band maximum (VBM).
The peaks in the vicinity of VBM are the flat-bands of the O$_{2p_z}$-Cu$_{3d(3z^2-r^2)}$-O$_{2p_z}$, O$_{2p_z}$-Ag$_{4d(3z^2-r^2)}$-O$_{2p_z}$ and O$_{2p_z}$-Au$_{5d(3z^2-r^2)}$-O$_{2p_z}$ anti-bonding $\pi$-bands in the pseudo-two dimensional (2D) frustrated triangular-lattices in the 2D Delafossite structures.
The band-gap of the AlO$_2$ layer is very large (8$\sim$9 eV) and the strong $p$-$d$ hybridized valence bands of O-Cu-O ($\sim$4 eV), O-Ag-O ($\sim$5 eV) and O-Au-O ($\sim$6 eV) dumbbells are inserted into the large band-gap of the octahedral AlO$_2$ layers.
Therefore, the VBM is the anti-bonding states, so that the hole-doping increases the covalency leading to the small formation energy of $p$-type dopants, such as the Cu-vacancy and O-interstitials~\cite{Hamada_6}.}
\label{DOS}
\end{center}
\end{figure}

\begin{table}[h]
\begin{center}
\begin{tabular}{c|ccc}
\hline\hline
& CuAlO$_2$ & AgAlO$_2$ & AuAlO$_2$\\ \hline
a (\AA) &  2.859  &  2.895   &  2.913\\
c/a &  5.965  &  6.369   &  6.373\\
z &  0.1101 &  0.1148  &  0.1158\\
\hline\hline
\end{tabular}
\caption{Optimized cell parameters for CuAlO$_2$, AgAlO$_2$ and AuAlO$_2$.}
\label{cell}
\end{center}
\end{table}
Figure \ref{band} depicts the band structures of 2D Delafossite (a) CuAlO$_2$, (b) AgAlO$_2$ and (c) AuAlO$_2$.
Figure \ref{DOS} shows the total density of states (DOS) and partial DOS of non-doped 2D Delafossite (a) CuAlO$_2$, (b) AgAlO$_2$ and (c) AuAlO$_2$.
The optimized cell parameters (see Fig.\ \ref{crystal}) are shown in Table 1.
Present calculated result of CuAlO$_2$ by the ultra-soft pseudoptential method is quantitatively in good agreement with the previous calculations by FLAPW method~\cite{HKY_2}.
The band-gap of the octahedral AlO$_2$ layer is very large (8$\sim$9 eV), and the strong $p$-$d$ hybridized bands of O-Cu-O (total valence band width is $\sim$4 eV), O-Ag-O ($\sim$5 eV) and O-Au-O ($\sim$6 eV) dumbbells are inserted into the large band-gap of octahedral AlO$_2$ layers.
Therefore, the top of the valence band of CuAlO$_2$, AgAlO$_2$ and AuAlO$_2$ is anti-bonding $\pi$-band states ({\it this is very unusual band structures compared with the conventional insulators or semiconductors, where the valence bands and VBM are always bonding states}) and flat-band structures due to the 2D with the frustrated triangular-lattices in the O-Cu-O, O-Ag-O and O-Au-O dumbbells array, and has a small peak at the VBM in the DOS, which is mainly constructed by the anti-bong $\pi$-band of Cu$_{3d(3z^2-r^2)}$ and O$_{2p_z}$ orbitals in the partial DOS in Fig.\ \ref{DOS}(a).
The 2D stacked layered structures of the O-Cu-O, O-Ag-O and O-Au-O dumbbells and octahedral AlO$_2$ layers in 2D Delafossite structure (Fig.\ \ref{crystal}, just like a natural super-lattices) of CuAlO$_2$, AgAlO$_2$ and AuAlO$_2$ show the very flat-band caused by the O$_{2p_z}$-Cu$_{3d(3z^2-r^2)}$-O$_{2p_z}$, O$_{2p_z}$-Ag$_{4d(3z^2-r^2)}$-O$_{2p_z}$, and O$_{2p_z}$-Au$_{5d(3z^2-r^2)}$-O$_{2p_z}$ anti-bonding $\pi$-band in the pseudo-2D frustrated triangular-lattices.
Inside the dumbbell, due to the strong $p$-$d$ hybridization in the $\sigma$-bond between the Cu$_{3d(3z^2-r^2)}$ and O$_{2p_z}$ forms bonding and anti-bonding states in the $z$-direction.
Next step, we can distribute these bonding and anti-bonding orbitals in the 2D triangular lattices, so that these orbitals form very frustrated anti-bonding $\pi$-band, indicating the very flat-band structures at VBM.
In CuAlO$_2$, the first peak at the VBM in the DOS is anti-bonding state caused by the strong $p$-$d$ hybridization between Cu$_{3d(3z^2-r^2)}$ and O$_{2p_z}$, the second and third high peaks from the VBM are non-bonding states constructed by the Cu$_{3d(x^2-y^2)}$ and Cu$_{3d(xy)}$, and the fourth peak from the VBM is bonding state by the strong $p$-$d$ hybridization between Cu$_{3d(3z^2-r^2)}$ and O$_{2p_z}$.
Since the top of the valence band (flat-band) consists of the anti-bonding states and is occupied by electrons, one can easily make $p$-type doping ({\it holes at the VBM}) due to the small formation energy of the Cu-vacancy in the Cu-poor crystal growth condition or the Oxygen interstitials in the Oxygen-rich crystal growth conditions, as was discussed by Hamada and Katayama-Yoshida~\cite{Hamada_6}.

The calculated band-gaps ($E_g=1.83$ eV for CuAlO$_2$, $E_g=1.47$ eV for AgAlO$_2$ and $E_g=0.46$ eV for AuAlO$_2$) by LDA are smaller than the experimentally observed values due to the LDA errors, where the LDA underestimates the band-gap by the self-interaction.
In order to correct the band-gap by taking into account the self-interaction correction (SIC), we have calculated the electronic structure by pseudo-SIC method (PSIC-LDA), where $E_g=3.16$ eV for CuAlO$_2$, $E_g=2.43$ eV for AgAlO$_2$ and $E_g=1.21$ eV for AuAlO$_2$, and obtained better agreement than LDA with the experimental data~\cite{Nakanishi_12}, where $E_g=2.96$ eV for CuAlO$_2$.
The calculated band widths of the flat-bands caused by the O$_{2p_z}$-Cu$_{3d(3z^2-r^2)}$-O$_{2p_z}$, O$_{2p_z}$-Ag$_{4d(3z^2-r^2)}$-O$_{2p_z}$, and O$_{2p_z}$-Au$_{5d(3z^2-r^2)}$-O$_{2p_z}$ anti-bonding $\pi$-bands in the pseudo-2D frustrated triangular lattices are as follows: $W \approx$ 1.7 eV for CuAlO$_2$, $W \approx 2.8$ eV for AgAlO$_2$, and $W \approx 3.5$ eV for AuAlO$_2$ (see in Fig.~\ref{band}).
As will discussed in Sec.~\ref{sec:BCS_BEC}, in the theoretically proposed negative $U_{\rm eff}$ Hubbard model from the first principles, the relation between the $|U_{\rm eff}|$ and $W$ is essentially important for discriminating the superconductivity coupling regime to discuss the crossover from the weak-coupling BCS regime to the strong-coupling BEC regime.

\subsection{Design by STEP 1 (HOP): Find $T_{\rm c}$ and $S$-Wave Superconductors by Electron-Phonon Interactions from the First Principles}
Nakanishi and Katayama-Yoshida~\cite{Nakanishi_14} calculated the superconducting critical temperature ($T_c=50\ {\rm K}$ for CuAlO$_2$) by electron-phonon interaction, based on the Step 1 as we discussed in Sec.~\ref{sec:Strategies}, using the Allen-Dynes modified McMillan's formula~\cite{McMillan_31,Allen_32}.
According to this formula, $T_c$ is given by three parameters: the electron-phonon interaction $\lambda$, the logarithmic averaged phonon frequency $\omega_{\log}$, and the screened Coulomb interaction $\mu^*$.
$\lambda$ and $\omega_{\log}$ are obtained by {\it ab initio} calculations using the density functional perturbation theory.
As for $\mu^*$, we assume the value $\mu^* = 0.1$.
This value holds for weakly correlated materials due to the electronic structure of lightly hole-doped Cu$^+(3d^{10})$.
As mentioned above, CuAlO$_2$ has the O$_{2p_z}$-Cu$_{3d(3z^2-r^2)}$-O$_{2p_z}$ anti-bonding $\pi$-band electrons at the top of the valence band.
When we make the hole doping, the electrons which make the O-Cu-O anti-bonding bands are located at the Fermi level.
They have a strong interaction with the $A_1L_1$ phonon mode because their bonding directions are parallel to the oscillation direction of the $A_1L_1$ phonon mode.
Though the strong electron-phonon interaction, the O-Cu-O bonding of the 2D Delafossite structure is stable even under a high pressure~\cite{Nakanishi_12}.

As $d$-wave function extends from $3d\rightarrow4d\rightarrow5d$ in order, the $d$-band width becomes broader, leading to weaker covalent bonding.
Therefore this low covalency of AuAlO$_2$ is the reason of low electron-phonon interaction ($\lambda \sim 0.931$ and $T_c \sim 50\ {\rm K}$ for CuAlO$_2$ in the case of optimized hole-doping, $\lambda \sim 0.9$ and $T_c \sim 40\ {\rm K}$ for AgAlO$_2$, and $\lambda \sim 0.45$ and $T_c \sim 3\ {\rm K}$ for AuAlO$_2$~\cite{Nakanishi_14,Nakanishi_15}) caused by the pseudo-2D nested Fermi surfaces originating from the flat-band of O-Cu-O, O-Ag-O and O-Au-O dumbbells, if we can do ideal hole-doping without disturbing the flat-band structure in the two-dimensionally stacked O-Cu-O, O-Ag-O and O-Au-O dumbbells in the CuAlO$_2$, AgAlO$_2$ and AuAlO$_2$.
It is also clearly seen in Fig.\ \ref{band} that the flatness of the top of the valence band decreases from CuAlO$_2$ to AuAlO$_2$ with reducing the covalency ($p$-$d$ hybridization).
We calculated the chemical trend of $T_c$ by the electron-phonon interactions of the hole-doped pseudo-2D Delafossite CuAlO$_2$, AgAlO$_2$ and AuAlO$_2$.
Calculated $T_c$ values are about 50 K, 40 K and 3 K for CuAlO$_2$, AgAlO$_2$ and AuAlO$_2$, respectively, at a maximum.
The low $T_c$ of AuAlO$_2$ is attributed to the low covalency and heavy atomic mass of Au.

There is a strong possibility that the crystal of the hole-doped CuAlO$_2$ becomes more stable upon the doping because the holes go to the anti-bonding state to stabilize the crystal structures by introducing the more covalency.
Generally, if the conventional materials are heavily hole-doped, the doped holes always go to the bonding state in the conventional insulator or semiconductors, and then the total electron-phonon interaction decreases because the number of electrons, which have a strong interaction, decreases.
However, since the top of the valence band of the 2D Delafossite CuAlO$_2$ is constructed by the Cu $3d_{3z^2-r^2}$ and O$_{2p_z}$ anti-bonding $\pi$-band, the doped holes which go to the anti-bonding states of CuAlO$_2$ make the O-Cu-O coupling more strong.
Therefore, when the density of the holes increases from 0.2 to 0.3, $\lambda$ and $\omega_{\log}$ does not change much ($\lambda = 0.901 \rightarrow 0.931$, $\omega_{\log} = 808\ {\rm K} \rightarrow 789\ {\rm K}$)~\cite{Nakanishi_14}.

\subsection{Design by STEP 2 (STEP): Design of Itinerant and Attractive Fermion System by Purely Electronic Negative U$_{\rm eff}$ Mechanism}
Based on the total energy calculation of the hole-doped 2D Delafossite structures of CuAlO$_2$, AgAlO$_2$ and AuAlO$_2$, we investigate the $U_{\rm eff}$ value: $U_{\rm eff} = E(N+1)+E(N-1)-2E(N)$, where $E(N)$ is the total energy of $N$ electron system.
The total energy calculations of the hole-doped Delafossite materials are performed for the different charge states of $E(N-1)$, $E(N)$ and $E(N+1)$ electron systems for the hole-doped 2D CuAlO$_2$, AgAlO$_2$ and AuAlO$_2$ by decreasing the number of electrons in the unit cell with keeping the charge neutrality by adding the jellium background charge in order to satisfy the real experimental condition upon the chemical doping from the charge reservoir.
In the real systems, we can dope the electron (hole) to the system by taking electron (hole) from the charge reservoir and put it to the Fermi level in the unit cell with keeping the hole (electron) in the charge reservoir with uniform jellium background.
This always satisfies charge neutrality in the unit cell, since the real chemical doping by acceptors or donors never charge up the host crystals upon the doping.

\begin{figure}[t]
\begin{center}
\includegraphics[height=8cm,clip]{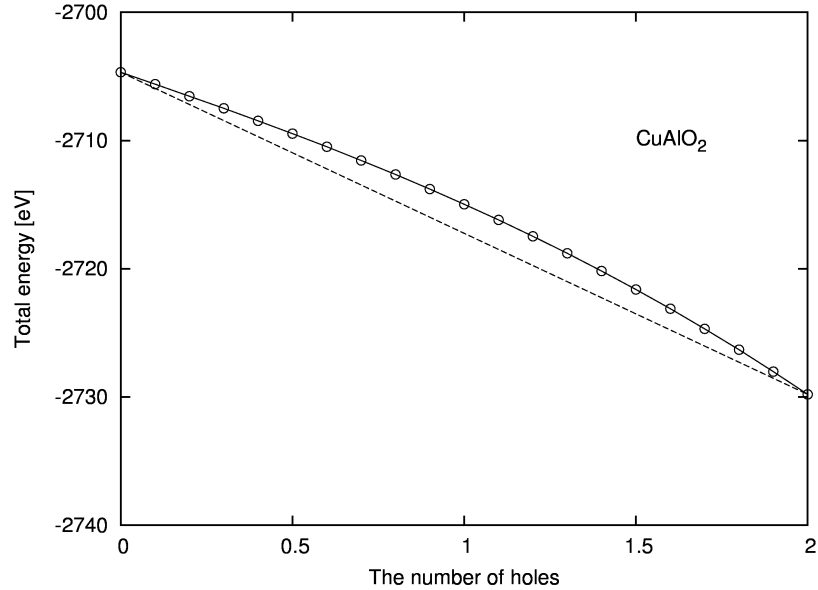}
\caption{Total energy of the 2D Delafossite structure of CuAlO$_2$ as a function of the number of the doped holes by LDA.
The total energy shows a convexity upward indicating negative $U_{\rm eff}$:  $U^{\rm (i)}_{\rm eff} =E(N+1)+E(N-1)-2E(N)= -4.53$ eV and $U^{\rm (ii)}_{\rm eff} \simeq \frac{\partial^2E(n)}{\partial n^2}|_{n=N}=-4.54$ eV. 
The dashed line is the guide for eyes.
The large negative $U_{\rm eff}$ leads to the very large attractive Fermion system and Bose-Einstein condensation (BEC)~\cite{Nozieres_23} by forming spin singlet $(S=0)$ bound states (composite Bosons) in the case of $|U_{\rm eff} = -4.53\ {\rm eV}| > W$ (1.7 eV, see Fig.~\ref{band}(a)) for hole-doped CuAlO$_2$, where $W$ is the band width.}
\label{Etot_Cu}
\end{center}
\end{figure}

\begin{figure}[t]
\begin{center}
\includegraphics[height=8cm,clip]{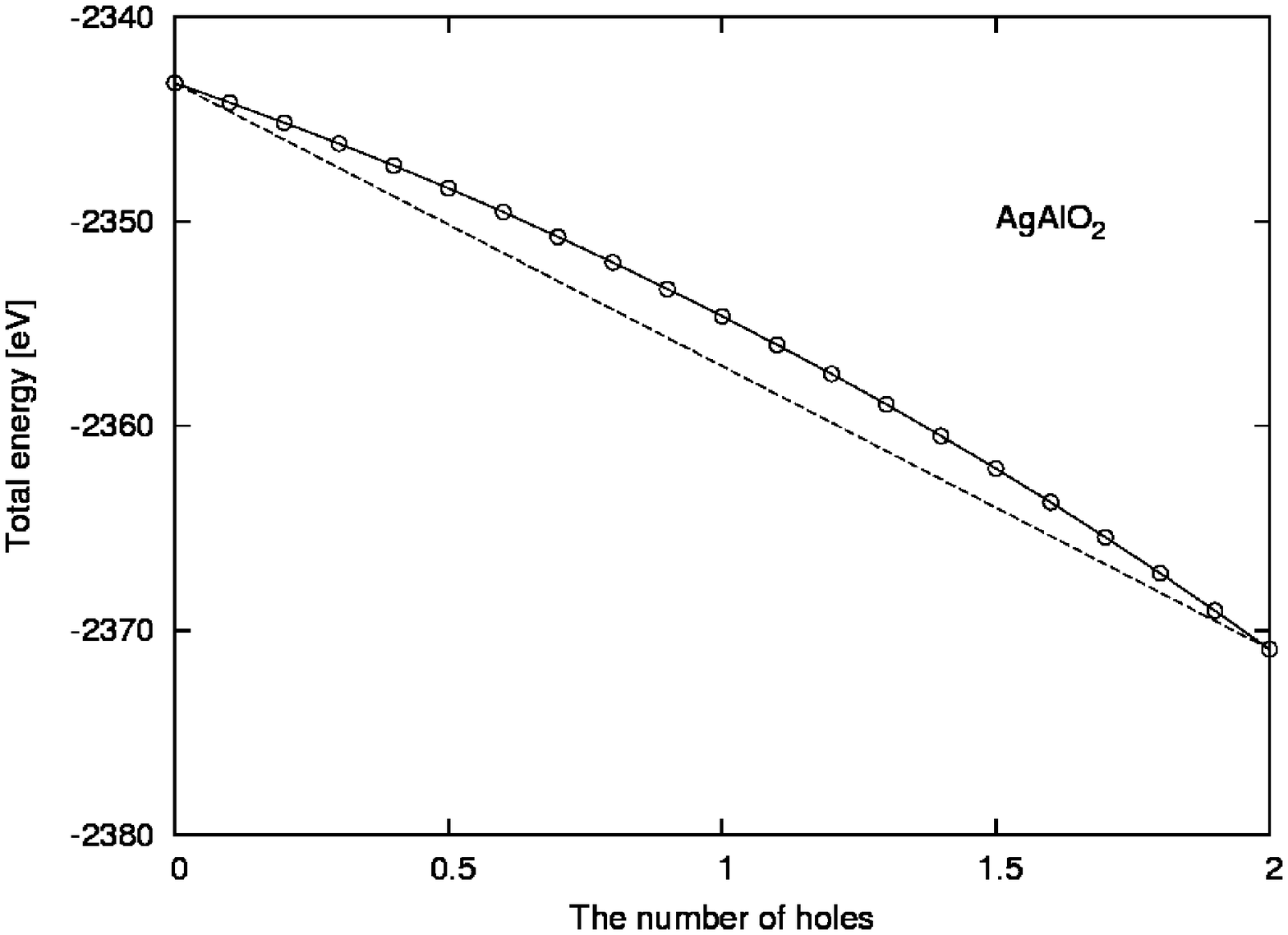}
\caption{Total energy of the 2D Delafossite structure of AgAlO$_2$ as a function of the number of the doped holes by LDA.
The total energy shows a convexity upward indicating negative $U_{\rm eff}$: $U^{\rm (i)}_{\rm eff} =E(N+1)+E(N-1)-2E(N)= -4.88$ eV and $U^{\rm (ii)}_{\rm eff} \simeq \frac{\partial^2E(n)}{\partial n^2}|_{n=N}=-4.88$ eV.
The dashed line is the guide for eyes.
The large negative $U_{\rm eff}$ leads to the very large attractive Fermion system and the crossover between the Bardeen-Cooper-Schrieffer (BCS) and Bose-Einstein condensation (BEC) in the case of $|U_{\rm eff} = -4.88\ {\rm eV} | \approx$ band width $W$ (2.8 eV, see Fig.~\ref{band}(b)) for the hole-doped 2D Delafossite of AgAlO$_2$.}
\label{Etot_Ag}
\end{center}
\end{figure}

\begin{figure}[t]
\begin{center}
\includegraphics[height=8cm,clip]{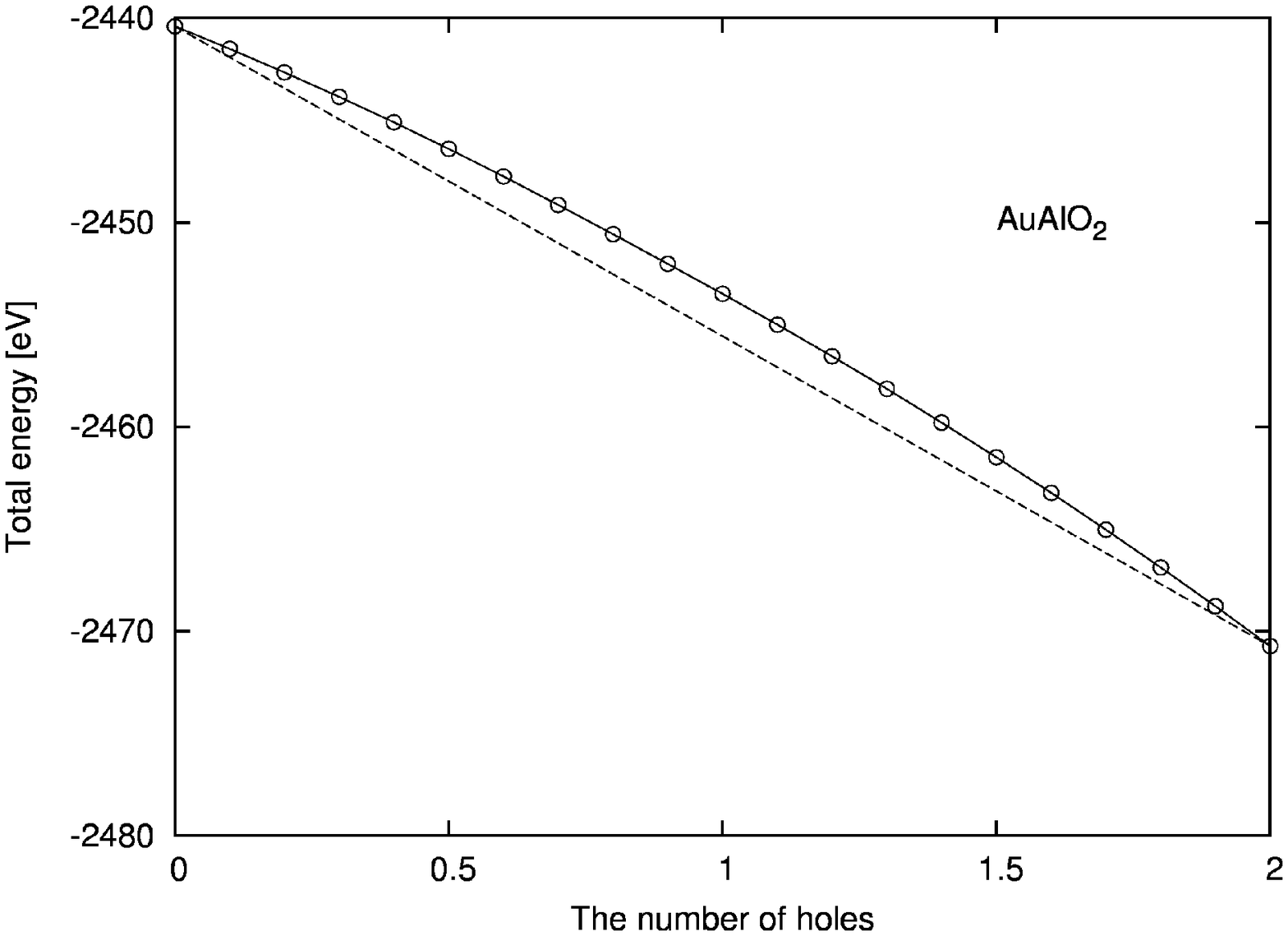}
\caption{Total energy of the 2D Delafossite of AuAlO$2$ as a function of the number of the doped holes by LDA.
The total energy shows a convexity upward indicating negative $U_{\rm eff}$: $U^{\rm (i)}_{\rm eff} =E(N+1)+E(N-1)-2E(N) = -4.14$ eV and $U^{\rm (ii)}_{\rm eff}\simeq \frac{\partial^2E(n)}{\partial n^2}|_{n=N}=-4.14$ eV.
The dashed line is the guide for eyes.
The large negative $U_{\rm eff}$ leads to the very large attractive Fermion system and the crossover between the Bardeen-Cooper-Schrieffer (BCS) and Bose-Einstein condensation (BEC) in the case of $|U_{\rm eff} = -4.14\ {\rm eV}| \approx$ band width $W$ (3.5 eV, see Fig.~\ref{band}(c)) for the hole-doped 2D Delafossite of AuAlO$_2$.}
\label{Etot_Au}
\end{center}
\end{figure}

Figures~\ref{Etot_Cu}, \ref{Etot_Ag} and \ref{Etot_Au} show the calculated total energies of the hole-doped 2D CuAlO$_2$, AgAlO$_2$ and AuAlO$_2$ as a function of the number of doped holes.
The total energies of these systems decrease rapidly with increasing the doped holes.
This is because the additional holes go to the VBM (anti-bonding states), so that the covalent stabilization increases.
All of the cases, the total energies have strong convexity upward as a function of doped holes, indicating the negative $U_{\rm eff}$.
Present results are consistent with the unified general design rules (see Sec.~\ref{sec:General_rule}) for the negative $U_{\rm eff}$ and the attractive Fermion system.
This unusual physical property is due to the special electronic structure of 2D Delafossite CuAlO$_2$, AgAlO$_2$ and AuAlO$_2$, in which the VBM is the anti-bonding state caused by the strong $p$-$d$ hybridizations between the Cu$_{3d}$-O$_{2p}$, Ag$_{4d}$-O$_{2p}$ and  Au$_{5d}$-O$_{2p}$ in the layered O-Cu-O, O-Ag-O and O-Au-O dumbbells in the natural super-lattices.
In the LDA calculations, the $U_{\rm eff}$ valued are -4.53, -4.88 and -4.14eV for CuAlO$_2$, AgAlO$_2$ and AuAlO$_2$, respectively.
The large negative $U_{\rm eff}$ upon the hole doping originates from the charge-excitation mechanism in the stability of the closed shells of Cu$^+(3d^{10})$, Ag$^+(4d^{10})$, Au$^+(5d^{10})$ and O$^{2-}(2p^6)$.
In the ground states of the closed shells of Cu$^+(3d^{10})$, Ag$^+(4d^{10})$, Au$^+(5d^{10})$ and O$^{2-}(2p^6)$ electronic configurations, the first excited states are Cu$^+(3d^94s^1)$, Ag$^+(4d^95s^1)$, Au$^+(5d^96s^1)$ and O$^{2-}(2p^53s^1)$, respectively. The ground state and first excited state are very close in each other; therefore, the two states repel so strongly in the second order perturbation, resulting in the stabilization of the ground state of the closed shell.
By the hole doping, the hole-doped Cu$^{2+}(3d^9)$, Ag$^{2+}(4d^9)$, Au$^{2+}(5d^9)$ and O$^{2-}(2p^6)$ become unstable relative to the ground states in the closed shells of Cu$^+(3d^{10})$, Ag$^+(4d^{10})$, Au$^+(5d^{10})$ and O$^{2-}(2p^6)$.
Due to the above reasons, one can obtain the negative $U_{\rm eff}$ for Cu$^{2+}(3d^9)$, Ag$^{2+}(4d^9)$, Au$^{2+}(5d^9)$ and O$^-(2p^5)$ electronic configurations caused by the stability of closed shells in Cu$^+(3d^{10})$, Ag$^+(4d^{10})$, Au$^+(5d^{10})$ and O$^{2-}(2p^6)$.

In order to check the SIC in the LDA, we performed PSIC-LDA calculations~\cite{Nakanishi_13}. 
The PSIC-LDA lead to the following negative $U_{\rm eff}$: CuAlO$_2$ $(U_{\rm eff} = -4.20\ {\rm eV})$, AgAlO$_2$ $(U_{\rm eff} = -4.55\ {\rm eV})$ and AuAlO$_2$ $(U_{\rm eff} = -3.57\ {\rm eV})$.
Since the electronic structures by the PSIC-LDA show more localized 3$d$-, 4$d$- and 5$d$-orbitals than the LDA, the PSIC-LDA gives more weak chemical bonding states.
The microscopic mechanism of the negative $U_{\rm eff}$ is attribute to the charge-excitation-induced mechanism in the stability of closed shells.
Therefore, one can expect that the PSIC-LDA gives smaller value on $|U_{\rm eff}|$ than that of the LDA. 
We also calculate the $U_{\rm eff}$ by the LDA and PSIC-LDA based on two definitions: (i) $U^{\rm (i)}_{\rm eff} = E(N+1) + E(N-1) -2E(N)$ and (ii) $U^{\rm (ii)}_{\rm eff} = E(N+1) + E(N-1) -2E(N) \simeq \partial^2E(n)/\partial n^2 |_{n=N}$.
The obtained $U_{\rm eff}$ values by (i) and (ii) are reasonable agreement each other, {\it i.e.,} for the LDA calculations, CuAlO$_2$ ($U^{\rm (i)}_{\rm eff} = -4.53\ {\rm eV}$, $U^{\rm (ii)}_{\rm eff} = -4.54\ {\rm eV})$, AgAlO$_2$ ($U^{\rm (i)}_{\rm eff} = -4.88\ {\rm eV}$, $U^{\rm (ii)}_{\rm eff} = -4.88\ {\rm eV}$) and AuAlO$2$ ($U^{\rm (i)}_{\rm eff} = -4.14\ {\rm eV}$, $U^{\rm (ii)}_{\rm eff} = -4.14\ {\rm eV}$), and for the PSIC-LDA calculations, CuAlO$_2$ ($U^{\rm (i)}_{\rm eff}=-4.20\ {\rm eV}$, $U^{\rm (ii)}_{\rm eff} = -4.19\ {\rm eV}$), AgAlO$_2$ ($U^{\rm (i)}_{\rm eff}=-4.55\ {\rm eV}$, $U^{\rm (ii)}_{\rm eff} = -4.54\ {\rm eV}$) and AuAlO$_2$ ($U^{\rm (i)}_{\rm eff}=-3.55\ {\rm eV}$, $U^{\rm (ii)}_{\rm eff} = -4.14\ {\rm eV}$).


In Table \ref{CuAlO2_ele} and \ref{CuFeS2_ele}, the projected partial charges of the Cu-$3d$ orbitals upon the hole doping for the 2D Delafossite CuAlO$_2$ and 3D Chalcopyrite CuFeS$_2$ are shown for the comparison.
The number of the local $3d$ electron charge at the Cu-site does not change very much due to the strong $p$-$d$ hybridization (Haldane and Anderson mechanism~\cite{Haldane_33,HKY_34}) upon the doping.
The change of the Cu-$3d$ electron $(\Delta Q)$ upon the hole-doping is $\Delta Q \approx 0.3$, and this value for the 2D Delafossite CuAlO$_2$ is ten times larger than the 3D hole-doped Chalcopyrite CuFeS$_2$ $(\Delta Q \approx 0.04)$.
When one hole is doped to CuAlO$_2$ (CuFeS$_2$), the doped 0.3 (0.04) holes are located at the Cu-$3d$ site.
However, the remaining doped 0.7 (0.96) holes are delocalized through the valence band by the strong $p$-$d$ hybridization.
Therefore, one can expect the strong screening of the $U_{\rm eff}$ through the $p$-$d$ hybridization.
According to the Haldane and Anderson mechanism~\cite{Haldane_33,HKY_34}, the reduction of the $U_{\rm eff}$ from the bare $U_0$ of Cu-atom by the screening due to the strong $p$-$d$ hybridization is approximately written as $|U_{\rm eff}| \approx 1/2(\Delta Q)^2|U_0|$.
The $U_{\rm eff}$ value of the hole-doped 3D Chalcopyrite CuFeS$_2$ calculated by {\it ab initio} calculations is $-0.44$ eV~\cite{Fukushima_19}, which satisfies the above criterion, as shown in Figs.\ \ref{CuFeS2_band} and \ref{CuFeS2_energy}.
The large $|U_{\rm eff}|$ value of CuAlO$_2$ compared to the case of CuFeS$_2$ is due to the poor screening in the 2D $p$-$d$ hybridization.
This is why the $|U_{\rm eff}|$ value of CuAlO$_2$ should be 10 times larger than that of CuFeS$_2$.
Although the negative $U_{\rm eff}$ is insensitive to the Cu$_{3d}$-, Ag$_{4d}$- and Au$_{5d}$-orbitals in the chemical trend, the dimensionality of the 2D Delafossite CuAlO$_2$ and 3D Chalcopyrite CuFeS$_2$ strongly affects the negative $U_{\rm eff}$ because of the strong dimensional dependence of the $p$-$d$ hybridization ({\it $\pi$-bond and two-fold coordination in Delafossite vs. $\sigma$-bond and four-fold coordination in Chalcopyrite}).
\begin{table}[h]
\begin{center}
\begin{tabular}{c|ccc}
\hline\hline
Number of holes & Cu-$3d$ & Cu-$4s$ & Cu-$4p$\\ \hline
 0        &       9.628     &   0.465    &    0.787\\
1          &     9.304      &  0.358      &  0.750\\
2          &     9.126    &    0.285      &  0.671\\
\hline\hline
\end{tabular}
\caption{Number of the electrons projected to the Cu-$3d$, Cu-$4s$ and Cu-$4p$ orbital components upon the hole doping in 2D Delafossite CuAlO$_2$.}
\label{CuAlO2_ele}
\end{center}
\end{table}
\begin{table}[h]
\begin{center}
\begin{tabular}{c|ccc}
\hline\hline
Number of holes & Cu-$3d$ & Cu-$4s$ & Cu-$4p$\\ \hline
 0        &       9.221     &   0.376    &    0.367\\
1          &     9.182      &  0.406      &  0.419\\
2          &     9.152    &    0.436      &  0.469\\
\hline\hline
\end{tabular}
\caption{Number of the electrons projected to the Cu-$3d$, Cu-$4s$ and Cu-$4p$ orbital components upon the hole doping in 3D Chalcopyrite CuFeS$_2$.}
\label{CuFeS2_ele}
\end{center}
\end{table}


Here, we should discuss an important duality of the electronic structures in hole-doped 2D Delafossite CuAlO$_2$, AgAlO$_2$ and AuAlO$_2$ upon the hole doping, by choosing CuAlO$_2$ as a example.
One is that the stability of the crystal structure is happened by the increased covalency in the chemical bonds of the O-Cu-O dumbbells, where the doped-holes go to the anti-bonding $\pi$-band at VBM by increasing the covalency with localized hyper-deep bonding electron in the deep bonding states far below the VBM.
The other is that the attractive electron-electron interaction in the itinerant Fermion system in the VBM is caused by the charge-excitation-induced negative $U_{\rm eff}$ upon the hole doping, where the charge-excitation or charge-fluctuation between the Cu and O in the O-Cu-O dumbbells dominates the negative $U_{\rm eff}$ mechanism, with keeping the crystal structure.
As was discussed before, the energy difference between the ground state of Cu$^+(3d^{10})$ electronic configurations and the first excited state of Cu$^+(3d^94s^1)$ electron configurations are very small, so that due to the strong charge excitation mechanism in the anti-bonding (delocalized) $\pi$-band, the ground state of Cu$^+(3d^{10})$ electronic configuration is stabilized.
However, in the case of the hole-doped CuAlO$_2$, the energy difference between the ground state of Cu$^{2+}(3d^9)$ electronic configuration and the first excited state of Cu$^{2+}(3d^84s^1)$ electron configuration are large compared with the case of the un-doped CuAlO$_2$.
Therefore, the hole-doped Cu$^{2+}(3d^9)$ electronic configuration is destabilized by the weak charge-excitation mechanism.
As a result, the strong convexity upward in the total energy as a function of the doped holes is obtained as shown in Figs.\ \ref{Etot_Cu}, \ref{Etot_Ag} and \ref{Etot_Au}.

\subsection{Design by STEP 3 (JUMP) : Mapping on the Negative $U_{\rm eff}$ Hubbard Model, Phase Diagram, and T$_c$}
\begin{figure}[t]
\begin{center}
\includegraphics[width=8cm,clip]{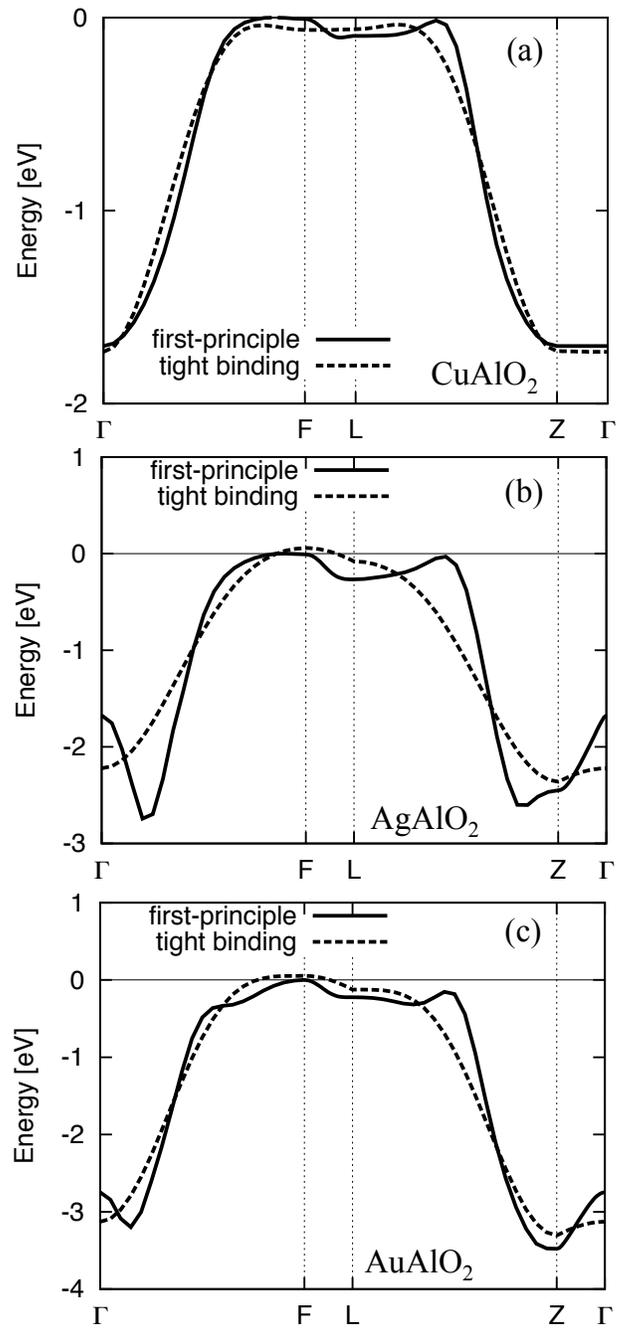}
\caption{Tight binding fitting for the highest valence bands of the anti-bonding $\pi$-bands mainly constructed by Cu$_{3d(3z^2-r^2)}$, Ag$_{4d(3z^2-r^2)}$ and Au$_{5d(3z^2-r^2)}$ orbitals in 2D Delafossite of (a) CuAlO$_2$, (b) AgAlO$_2$ and (c) AuAlO$_2$ by the LDA calculations and eq.\ (\ref{eq:fit})}
\label{fit}
\end{center}
\end{figure}

The calculated $U_{\rm eff}$ values for the hole-doped CuAlO$_2$, AgAlO$_2$ and AuAlO$_2$ indicate that these systems are very attractive Fermion systems with large negative $U_{\rm eff}$.
Based on {\it ab initio} electronic structure calculation of CuAlO$_2$, one can project the flat-band from the highest valence band of CuAlO$_2$ corresponding to a small peak in the DOS (see Fig.~\ref{DOS}(a) by the tight-binding band model.
This 2D flat-band is caused by the O$_{2p_z}$-Cu$_{3d(3z^2-r^2)}$-O$_{2p_z}$ anti-bonding $\pi$-band in the pseudo-2D frustrated triangular lattices in the Delafossite structure.
Using the calculated negative $U_{\rm eff}$ in CuAlO$_2$ and the fitted tight-binding band structure on the basis of {\it ab initio} electronic structure calculation, we can make mapping on the negative $U_{\rm eff}$ Hubbard model by choosing the single band, {\it i.e.,} the anti-bonding $\pi$-band at VBM.
Therefore, the phase diagram and $T_c$ are quantitatively predicted by the quantum Monte Carlo simulations without experimental data and empirical parameters~\cite{Scalettar_20,Moreo_21}.
When only the first nearest neighbor hopping is considered, the highest anti-bonding valence bands (the anti-bonding $\pi$-bands) of CuAlO$_2$, AgAlO$_2$ and AuAlO$_2$ are represented as follows:
\begin{eqnarray} \label{eq:fit}
\epsilon (k_x,k_y,k_z)&=&2t_1\cos(k_zd_{\rm A-B})+2t_2\cos(k_xd_{\rm A-A})\nonumber\\
&+&4t_3\cos(\frac{k_xd_{\rm A-A}}{2})\cos(\frac{\sqrt{3}k_yd_{\rm A-A}}{2})+\epsilon_0.
\end{eqnarray}
Here, $d_{\rm A-B}$ is the distance between A atom (A=Cu, Ag and Au) and B (B=O) atom, $t_i$ is the hopping parameter.
Table\ \ref{table_fit} and Fig.\ \ref{fit} show the results of the fitting parameters obtained from {\it ab initio} calculations and the fitted band structure of the anti-bonding $\pi$-band, respectively. 
\begin{table}[h]
\begin{center}
\begin{tabular}{c|cccccc}
\hline\hline
 & $d_{\rm A-A}$ (${\rm \AA}$) & $d_{\rm A-B}$ (${\rm \AA}$) & $\epsilon_0$ (eV) & $t_1$ (eV) & $t_2$ (eV) & $t_3$ (eV)\\ \hline
CuAlO$_2$ & $1.880$ & $2.863$ & $8.2875$ & $-0.0041$ & $-0.2841$ & $-0.1329$\\ 
AgAlO$_2$ & $2.117$ & $2.895$ & $7.8688$ & $0.1306$ & $-0.4963$ & $-0.0739$\\ 
AuAlO$_2$ & $2.151$ & $2.913$ & $9.2396$ & $0.1650$ & $-0.5965$ & $-0.1988$\\ 
\hline\hline
\end{tabular}
\caption{Fitting parameters of the tight-binding band structure of the highest valence anti-bonding valence band by using eq.\ (\ref{eq:fit}).
$d_{\rm A-B}$ represents the distance between A atom (A=Cu, Ag and Au) and B atom (B=O).}
\label{table_fit}
\end{center}
\end{table}

As mentioned in Sec.\ \ref{sec:Strategies}, the behavior of $T_{\rm c}$ depends on the calculated values of $|U_{\rm eff}|$ and $W$ in the negative $U_{\rm eff}$ Hubbard model: (1) the BCS weak coupling regime $(|U_{\rm eff}| < W)$, (2) the BCS/BEC crossover regime $(|U_{\rm eff}| \approx W)$ and (3) the BEC strong coupling regime $(|U_{\rm eff}| > W)$.
The phase diagram of the negative $U_{\rm eff}$ Hubbard model for 3D system was obtained by Dagotto {\it et al.}~\cite{Dagotto_35,Dagotto_36} based on the connection between the negative $U_{\rm eff}$ Hubbard model and lattice gage theories.
Therefore, once the negative $U_{\rm eff}$ Hubbard model is calculated, one can see the phase diagram and $T_c$ according to the above researches.
Although our proposed single-band negative $U_{\rm eff}$ Hubbard model is based on the projection to the highest energy band in the valence band and $U_{\rm eff}<0$, we can easily extend to the multi-band model.
The negative $U_{\rm eff}$ Hubbard model with the single-band includes the only attractive Fermion $(U_{\rm eff}<0)$ and itinerant band near VBM.
However, in the back ground, based on the {\it ab initio} electronic structure calculations, we should notice that the holes doped to the anti-bonding valence band stabilize the crystal structures with the strong covalency.

\section{\label{sec:BCS_BEC}BCS Superconductor and BEC in the Negative $U_{\rm eff}$ Hubbard Model\protect\\}

\begin{figure}[t]
\begin{center}
\includegraphics[width=8cm,clip]{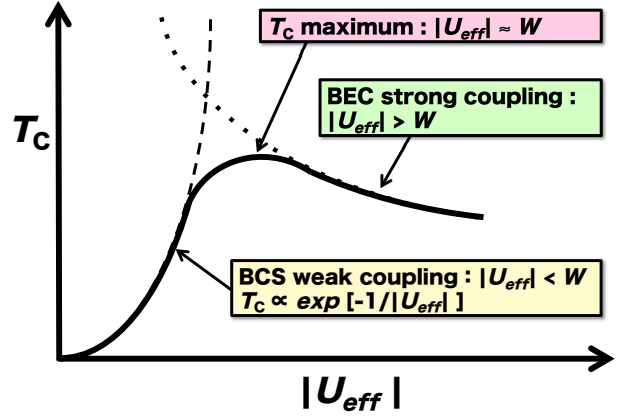}
\caption{Schematic description of $T_c$ vs. $|U_{\rm eff}|$ of the negative $U_{\rm eff}$ Hubbard model.
The negative $U_{\rm eff}$ Hubbard model by Nozi\'{e}res and Schmitt-Rink~\cite{Nozieres_23} indicates the evolution from weak- to strong-coupling superconductivity is continuous between Bardeen-Cooper-Schrieffer (BCS) and Bose-Einstein condensation (BEC) with increasing the $|U_{\rm eff}|$.
If the attractive interaction $|U_{\rm eff}|$ is smaller compared with the band width $(W)$ $(|U_{\rm eff}|<W)$, this system undergoes a superconducting instability at low temperature by forming Cooper pairs with a spin-singlet state $(S=0)$ according to the conventional BCS picture ({\it weak coupling regime}).
If the attractive interaction $|U_{\rm eff}|$ is larger compared with $W$ $(|U_{\rm eff}|>W )$, a bound state of a spin-singlet pair (composite Bosons) becomes possible and undergo to BEC ({\it strong coupling regime}).
$T_c$ increases exponentially $(T_c \propto \exp [-1/ |U_{\rm eff}| ] )$ in the BCS weak coupling regime $(|U_{\rm eff} = -0.44\ {\rm eV}| < W(\sim 2\ {\rm eV}))$ in the case of hole-doped 3D Chalcopyrite CuFeS$_2$, go through a maximum when $|U_{\rm eff} = -4.88\ {\rm eV}$ and $-4.14\ {\rm eV}| \approx$ band width $W$ (3 eV and 4 eV) in the case of hole-doped 2D Delafossite AgAlO$_2$ and AuAlO$_2$, and $T_C$ decreases with decreasing $U_{\rm eff}$ in the strong coupling regime, here $|U_{\rm eff} = -4.53\ {\rm eV}| > W (1.7\ {\rm eV)}$ in the case of hole-doped 2D Delafossite CuAlO$_2$.}
\label{N-SR}
\end{center}
\end{figure}

In 1985, for a negative $U_{\rm eff}$ Hubbard model, the relation between the BEC and BCS superconductivity was proposed by Nozi\'{e}res and Schmitt-Rink~\cite{Nozieres_23} (N-SR theory), as shown in Fig.~\ref{N-SR}.
In the N-SR theory, if the attractive interaction $|U_{\rm eff}|$ of Fermion by the negative $U_{\rm eff}$ is smaller than $W$ $(|U_{\rm eff}| < W)$, this system undergoes a superconducting instability at a low temperature by forming Cooper pairs with a spin-singlet state $(S=0)$ according to the conventional BCS picture ({\it weak coupling regime}).
If the attractive interaction $|U_{\rm eff}|$ of Fermion is larger than $W$ $(|U_{\rm eff}| > W )$, a bound state of a spin-singlet pair (composite Bosons) becomes possible, leading to the BEC ({\it strong coupling regime}).
They showed that the evolution from the weak- to strong-coupling superconductivity is smooth and seamless, and only the crossover can be seen between the BCS and BEC with increasing $|U_{\rm eff}|$ in the negative $U_{\rm eff}$ system.
The phase diagram of the negative $U_{\rm eff}$ Hubbard model for 3D system obtained by lattice gage theories~\cite{Dagotto_35, Dagotto_36} is consistent with the speculated phase diagram of the hole-doped CuFeS$_2$ by {\it ab initio} total energy calculation~\cite{Fukushima_19}.
Above phase diagram and crossover from the BCS to BEC is confirmed by the comparison between quantum Monte Carlo simulations and BCS/BEC crossover experiments by using the laser-cooled alkali-atom-gas with controlling the coupling constant (negative $U_{\rm eff}$) by Feshbach resonance effect.
In this BCS/BEC case, the negative $U_{\rm eff}$ is not sensitive to the details of the Fermi surfaces in the attractive Fermion systems.
However, it is not clear for electronic systems in condensed matter, if the microscopic mechanism of the negative $U_{\rm eff}$ is sensitive to the shape of the Fermi surfaces, just like a strong electron-phonon interaction or spin-fluctuation mechanism.
Our negative effective $U_{\rm eff}$ in the hole-doped 2D Delafossite of CuAlO$_2$ ($U_{\rm eff} = -4.53$ eV), AgAlO$_2$ ($U_{\rm eff} = -4.88$ eV) and AuAlO$_2$ ($U_{\rm eff} = -4.14$ eV), caused by {\it the charge-excitation-induced negative $U_{\rm eff}$ mechanisms}, is insensitive to the shape of the Fermi surfaces.
Therefore, we can expect that the $T_c$ should increase exponentially $(T_c \propto \exp [-1/ |U_{\rm eff}| ] )$ in the BCS weak coupling regime, and then the $T_{\rm c}$ should go through a maximum when $|U_{\rm eff}|  \approx$ $W$, and finally the $T_c$ should decrease with decreasing $U_{\rm eff}$ in the strong coupling regime, depending on the relation between $|U_{\rm eff}|$ and $W$, as was discussed by the N-SR theory~\cite{Nozieres_23}.

\begin{figure*}[t]
\begin{center}
\includegraphics[width=12cm,clip]{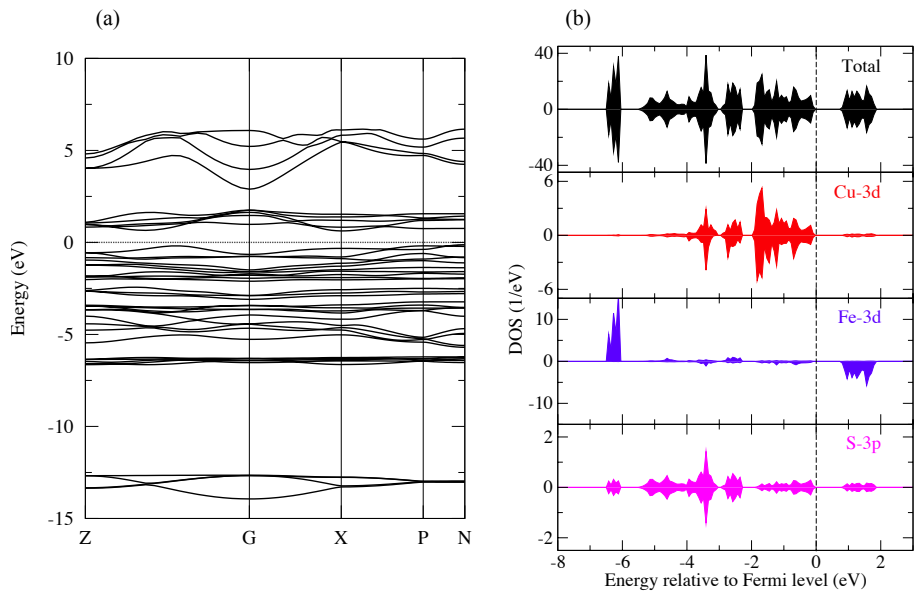}
\caption{(a) Spin-polarized band structure and (b) partial DOS of anti-ferromagnetic 3D Chalcopyrite CuFeS$_2$ by the GGA+$U$ method $(U - J = 4.0\ {\rm eV)}$.
The local magnetic moment of Fe$^{3+}(3d^5)$ is 3.8 $\mu_{\rm B}$, which is reduced by the strong covalency with strong $p$-$d$ hybridization.
The exchange-splitting of Fe$^{3+}(3d^5)$ is about 7 eV ($\approx$ Mott-Hubbard gap) with the high-spin ground state.
Strongly $p$-$d$ hybridized Cu-S bands (bonding, anti-bonding, and non-bonding states) is located inside the large Mott-Hubbard gap ($\sim$ 7 eV).
The band-gap $(E_g=0.6\ {\rm eV})$ of 3D Chalcopyrite CuFeS$_2$ is dominated by the charge-transfer energy from Cu to S, so called charge-transfer insulator, where [Mott-Hubbard correlation-gap energy] $>$ [Charge-transfer energy].}
\label{CuFeS2_band}
\end{center}
\end{figure*}

\begin{figure}[t]
\begin{center}
\includegraphics[width=8cm,clip]{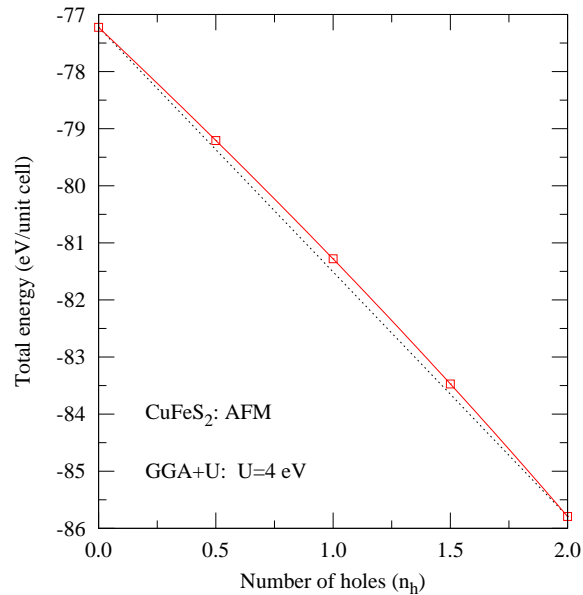}
\caption{Calculated total energy of 3D Chalcopyrite CuFeS$_2$ as a function of number of doped holes shows convexity upward indicating the negative effective $U_{\rm eff}=E(N+1)+E(N-1)-2E(N)<0$~\cite{Fukushima_19} for hole-doped CuFeS$_2$.
Calculated effective $U_{\rm eff}$ is negative and $U_{\rm eff} = -0.44\ {\rm eV}$.
The large negative $U_{\rm eff}$ $(U_{\rm eff} = -0.44\ {\rm eV})$ for the hole-doped system is originated by (1) the charge-excitation-induced negative $U_{\rm eff}$~\cite{HKY_17} mechanism of relatively unstable Cu$^{2+}(3d^9)$ caused by the energy gain of chemical bond in the closed shells of un-doped Cu$^+(3d^{10})$, and, (2) the exchange-correlation-induced negative $U_{\rm eff}$\cite{HKY_16} mechanism of hole-doped Fe$^{\rm 4+}(3d^4)$ caused by the energy gain of the Fe$^{3+}(3d^5)$ electronic configuration by Hund's rules relative to the hole-doped Fe$^{4+}(3d^4)$ in 3D Chalcopyrite of CuFeS$_2$.
$T_c$ increases exponentially $(T_c \propto \exp [-1/ |U_{\rm eff}| ] )$ in the Bardeen-Cooper-Schrieffer (BCS) weak coupling regime $(|U_{\rm eff} = -0.44\ {\rm eV}| < W (\sim2\ {\rm eV})$, see Fig.\ \ref{N-SR}) in the case of hole-doped 3D Chalcopyrite of CuFeS$_2$.}
\label{CuFeS2_energy}
\end{center}
\end{figure}

In the hole-doped 3D Chalcopyrite CuFeS$_2$, by a GGA+$U$ method ($U-J=4.0$ eV), Fukushima {\it et al.}~\cite{Fukushima_19} proposed that generally the $T_c$ should increase exponentially $(T_c \propto \exp [-1/|U_{\rm eff}| ] )$ in the BCS weak coupling regime ({\it as will discuss later, the hole-doped CuFeS$_2$ is located in this regime, because $|U_{\rm eff}| < W$, 
see Figs.\ \ref{CuFeS2_band} and \ref{CuFeS2_energy}}).
The predicted phase diagram of the anti-ferromagnetic, ferromagnetic and paramagnetic metal in the hole doped metallic CuFeS$_2$ shows a possibility of high-$T_c$ superconductor ($T_c \approx 1,000$ K, if $2{\Delta}/k_{\rm B}T_c =10$ by assuming a strong coupling regime and superconducting gap $\Delta \approx |U_{\rm eff}| \approx 5,000$ K) because of the large negative $U_{\rm eff}$ value.
Based on {\it ab initio} electronic structure calculations, we obtained the following $U_{\rm eff}$ and $W$: for the 3D Chalcopyrite CuFeS$_2$ ($U_{\rm eff} = -0.44$ eV, $W =2$ eV), 2D Delafossite CuAlO$_2$ ($U_{\rm eff} = -4.53$ eV, $W =1.7$ eV), AgAlO$_2$ ($U_{\rm eff} = -4.88$ eV,  $W =2.8$ eV) and AuAlO$_2$ ($U_{\rm eff} = -4.14$ eV, $W = 3.5$ eV).
In the framework of the N-SR theory, we can expect that the evolution from weak- to strong-coupling superconductivity is smooth and seamless.
Then, the only crossover can be seen between the BCS and BEC with increasing $|U_{\rm eff}|$.
In a theoretically proposed negative $U_{\rm eff}$ Hubbard model from first principles calculations, the relation between the $|U_{\rm eff}|$ and $W$ is essentially important to discriminate the BCS, BCS/BEC crossover and BEC regimes; (1) the BCS weak coupling regime $(|U_{\rm eff}| < W)$, (2) the BCS/BEC crossover regime $(|U_{\rm eff}| \approx W)$ and (3) the BEC strong coupling regime $(|U_{\rm eff}| > W)$.
By comparing the phase diagram in Fig.~\ref{N-SR} by the N-SR theory~\cite{Nozieres_23} and quantum monte simulation~\cite{Scalettar_20,Moreo_21,Dagotto_35,Dagotto_36} with the $|U_{\rm eff}|$ and $W$ obtained by first principles calculations, it is found that the hole-doped 3D Chalcopyrite CuFeS$_2$ corresponds to the BCS weak coupling regime $(T_c \propto \exp [-1/ |U_{\rm eff}| ] )$, $T_c$ goes through a maximum for the hole-doped 2D Delafossite of AgAlO$_2$ and AuAlO$_2$, and the hole-doped 2D Delafossite of CuAlO$_2$ is located at the strong coupling regime, where $T_c$ decreases with decreasing $U_{\rm eff}$, and $T_c \approx 1,000\sim2,000$ K, if $2\Delta/k_{\rm B}T_c =50\sim100$ by assuming a very strong coupling regime and superconducting gap $\Delta \approx |U_{\rm eff} | \approx 4.53\ {\rm eV} \approx 50,000$ K.

In order to confirm the unified general design rules for the negative $U_{\rm eff}$ and the qualitative discussion of the BCS and BEC crossover by the N-SR theory and {\it ab initio} electronic structure calculations of the hole-doped 2D Delafossite of CuAlO$_2$, AgAlO$_2$ and AuAlO$_2$ presented in this work, we would like to offer experimental verifications by developing new doping methods, such as co-doping~\cite{Yamamoto_38,Yamamoto_39,HKY_40,Bergqvist_41,Fujii_42} in thermal non-equilibrium crystal growth conditions, inhomogeneous spinodal nano-decomposition-based modulation doping~\cite{HKY_3} and doping by electric gating, and also offer quantum Monte Carlo simulations~\cite{Scalettar_20,Moreo_21,Dagotto_35,Dagotto_36} by computer simulation groups based on the present negative $U_{\rm eff}$ Hubbard model mapped from {\it ab initio} electronic structure calculations of the hole-doped 2D Delafossite of CuAlO$_2$, AgAlO$_2$ and AuAlO$_2$.
Compared with a strong electron-phonon interaction or spin-fluctuation mechanisms, our proposed mechanism of the super-high-$T_c$ by the charge-excitation-induced negative $U_{\rm eff}$ is insensitive to the shape of the Fermi surfaces.
Therefore, we need to check experimental verifications in the crossover between the BCS and BEC by controlling the $U_{\rm eff}$ and $T_c$ in the chemical trends, and the $T_c$ vs. $|U_{\rm eff}|$ and $W$ in the hole-doped 2D Delafossite of CuAlO$_2$, AgAlO$_2$, AuAlO$_2$ and hole-doped 3D Chalcopyrite of CuFeS$_2$.
We have a large possibility to realize the BEC in the attractive Fermion system with the negative $U_{\rm eff}$ electron system by growing the hole-doped 2D Delafossite of CuAlO$_2$, corresponding to the strong coupling regime in the BEC with the condition of $|U_{\rm eff}| > W$, where $U_{\rm eff} = -4.53\ {\rm eV}$ and the band width $W = 1.7\ {\rm eV}$.

Finally, we comment about recently-reported theoretical and experimental works on the relations between the missing oxidation states and attractive Fermion system with negative $U_{\rm eff}$, and on a possibility of the high-$T_{\rm c}$ superconductors by the negative $U_{\rm eff}$.
Strand studied, in a recent publication~\cite{Strand_48}, the valence-skipping and negative $U_{\rm eff}$ in $d$-bands from repulsive local Coulomb interaction point of view.
He proposed that the $U_{\rm eff}$ values become negative for the $d^1$, $d^4$, $d^5$, $d^6$ and $d^9$ electronic configurations in the $d$-band based on model calculations.
These results are consistent with (1) the charge-excitation-induced negative $U_{\rm eff}$ for $s^1$, $p^1$, $p^5$, $d^1$ and $d^9$ electronic configurations~\cite{HKY_17,Fukushima_19} caused by the stability of the chemical bond in the closed shells of $s^2$, $p^6$, $d^0$ and $d^{10}$ electronic configurations, and (2) the exchange-correlation-induced negative $U_{\rm eff}$ for $d^4$ and $d^6$ electronic configurations~\cite{HKY_16,Fukushima_19} due to the stability of $d^5$ electronic configuration in the Hund's rules with the high-spin states.
In a experimental work, Dorozdov {\it et al.} reported the conventional superconductivity with $T_c$ of around 190 K for $3{\rm H}_2{\rm S} \rightarrow 2{\rm SH}_3 + {\rm S}$ under a high pressure condition $P$ ($P> \sim 150$ GPa)~\cite{Dorozdov_49}.
In this case, according to the our general rules  for the missing oxidation states in Fig.~\ref{mo}, S$^{3+}$($s^2p^1$) is negative $U_{\rm eff}$.
By a chemical reaction, one can expect the charge fluctuation, $2{\rm S}^{3+}(s^2p^1) \rightarrow {\rm S}^{2+}(s^2p^2) + {\rm S}^{4+}(s^2p^0) + |U_{\rm eff}|$, which shows attractive electron-electron interactions in order to enhance the $T_c$ in the itinerant SH$_3$ Fermion system, as was discussed and shown in the present paper.

\section{\label{sec:Summary}Summary and Future Prospects\protect\\}
\begin{figure*}[t]
\begin{center}
\includegraphics[width=12cm,clip]{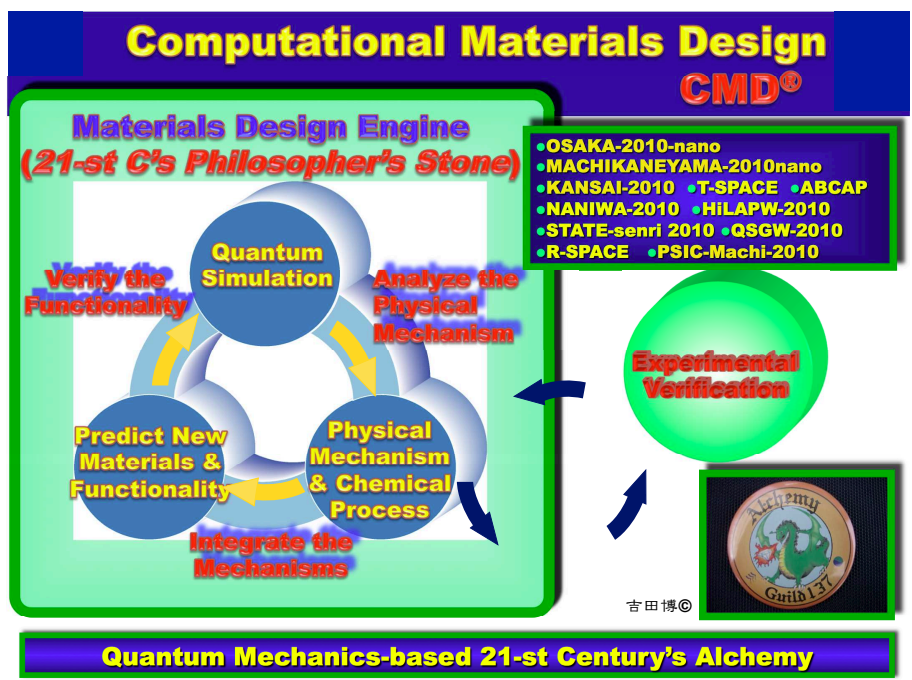}
\caption{Computational materials design (CMD$^\regmark$) system developed at Osaka University~\cite{HKY_37}.
Using the quantum simulation, we analyze a physical mechanism and chemical process, predict a new material and new functionality by integrating the physical mechanisms and chemical processes and verify the functionality.
By itinerating the materials design and experimental verification, we can design a new materials and new functionalities.}
\label{CMD}
\end{center}
\end{figure*}
In order to design the super-high-$T_c$ superconductors $(T_c > 1,000\ {\rm K})$ based on the unified general design rule of negative $U_{\rm eff}$ system by controlling the purely electronic and attractive Fermion mechanisms, we calculated the $U_{\rm eff}$ of the hole-doped 2D Delafossite CuAlO$_2$, AgAlO$_2$ and AuAlO$_2$ by the first principles with keeping the pseudo-two dimensional frustrated flat-band.
We found the large negative $U_{\rm eff}$ values caused by the charge-excitation-induced negative $U_{\rm eff}$ mechanism in the two dimensionally stacked O-Cu-O, O-Ag-O and O-Au-O dumbbells and AlO$_2$ layers, indicating the very attractive Fermion systems.
The calculated $U_{\rm eff}$ values are -4.53 eV, -4.88 eV and -4.14 eV for the hole-doped CuAlO$_2$, AgAlO$_2$ and AuAlO$_2$ by LDA, respectively, and -4.20 eV, -4.55 eV and -3.57 eV for the hole doped CuAlO$_2$, AgAlO$_2$ and AuAlO$_2$ by PSIC-LDA, respectively.
These negative $U_{\rm eff}$ values of the holed doped 2D Delafossite structure are almost ten times larger than the hole-doped 3D  Chalcopyrite CuFeS$_2$ ($U_{\rm eff} = -0.44$ eV) due to the poor screening in the 2D $p$-$d$ hybridization.
The mechanism of the large negative $U_{\rm eff}$ in the hole-doped 2D Delafossite is the charge-excitation-induced instability in $(N-1)$ electron system upon the hole-doping caused by the relatively stable chemical bond in the un-doped $(N)$ electron system in the closed shells of Cu$^+(3d^{10})$, Ag$^+(4d^{10})$ and Au$^+(5d^{10})$.
The chemical trend of the negative $U_{\rm eff}$ is insensitive to the Cu, Ag and Au because the origin of negative $U_{\rm eff}$ is chemical bond stability
However, it is very sensitive to the dimensionality between the 2D and 3D caused by the different $p$-$d$ hybridizations.

The negative $U_{\rm eff}$ Hubbard model shows the evolution from weak- to strong-coupling superconductivity is continuous between Bardeen-Cooper-Schrieffer (BCS) and Bose-Einstein condensation (BEC) with increasing the $|U_{\rm eff}|$ value~\cite{Scalettar_20,Moreo_21,Nozieres_23}.
We discussed the possibility of super-high-$T_c$ superconductors by our calculation results and N-SR theory.
$T_c$ should increase exponentially $(T_c \propto \exp [-1/ |U_{\rm eff}| ] )$ in the BCS weak coupling regime for the hole-doped 3D Chalcopyrite CuFeS$_2$ [see Figs.\ \ref{CuFeS2_band} and \ref{CuFeS2_energy}, and then $T_{\rm c}$ should go through a maximum when $|U_{\rm eff} = -4.88$ eV and $-4.14\ {\rm eV}| \approx$ band width $W$ (2.8 eV and 3.5 eV) for the hole-doped 2D Delafossite AgAlO$_2$ and AuAlO$_2$, and fainally $T_c$ should decrease with decreasing $U_{\rm eff}$ in the strong coupling regime, where $|U_{\rm eff} = -4.53\ {\rm eV}| > W$ (1.7 eV) for the hole-doped 2D Delafossite CuAlO$_2$.
We also proposed, in CuAlO$_2$, AgAlO$_2$ and AuAlO$_2$, the negative $U_{\rm eff}$ Hubbard model mapped from the {\it ab initio} electronic structure calculations, for the purpose of the future calculation of $T_c$ and phase diagram by the quantum Monte Carlo simulations.

Our society is now changing very rapidly from the industrial society to the knowledge-based society.
The social demands for the knowledge-based CMD$^\regmark$ increases dramatically in order to perform the efficient innovations in the research and developments of environment-friendly devices, energy-creation and energy-saving new materials, and the new-class of electronics devices including the spintronics, moltronics, and quantronics.
Especially for the post industrialized country, we do need to develop the knowledge-based CMD$^\regmark$ in order to adapt the change of the industrial structures or industrial hierarchy from the industrial society to the knowledge-based society~\cite{HKY_37,HKY_45,HKY_46}.
In this work, we showed that CMD$^\regmark$ for searching and designing of new super-high-$T_c$ superconductors based on the multi-scale simulations, combined with the {\it ab initio} electronic structure calculations and model-based theories, becomes more efficient and realistic for the future development of energy-related new materials in the 21-st century.
Figure.\ \ref{CMD} shows a CMD$^\regmark$ system developed at Osaka University~\cite{HKY_37,HKY_45,HKY_46}.
For the realization of more realistic and efficient knowledge-based CMD$^\regmark$ for high-efficiency energy conversion materials and energy saving materials, the following three steps are important for the CMD$^\regmark$ strategies.
STEP 1: analysis of the physical mechanism and chemical reactions mechanisms based on the quantum simulations, and proposal of the general rules for new materials and new functionalities.
STEP 2: design of the new functionality and new materials by integrating the physical and chemical mechanisms based on the general rules.
STEP 3: verification of the functionality by quantum simulation and obtained general rules for the new materials and new functionalities.
By itinerating the materials design and experimental verification, we can design a new materials and new functionalities.
Finally, the proposed functionality can be verified by quantum simulation.
If the newly designed materials do not meet our demands, the reason is analyzed and an alternative candidate as a new functional material is proposed through the analysis.
Using this circulation of design, realization and analysis in the CMD$^\regmark$ system, we can ultimately design the desired new functional materials.

So far, most of the experimental efforts to find new high-$T_c$ superconductors have been based on the 20-century's brute-force and trial-and-error methods with labor-intensive and low success-rate, or, accidental discovery methods with no-excellence, no-leadership and no-continuous innovations.
The new methodologies to design and create new materials, which we proposed in this paper, is one of the most important CMD$^\regmark$ for the future design and realization of super-high-$T_c$ superconductors.
We do need more work for the CMD$^\regmark$ and experimental realization to establish the future by CMD$^\regmark$, which is suitable for the 21-st century's alchemy based on the quantum theory and the multi-scale simulations by connecting the microscopic (nano-scale) quantum phase to the macroscopic and coherence quantum phases.

\ack
The authors acknowledge the financial support provided by the Japan Society for the Promotion of Science ``JSPS Core-to-Core'' Program ``Computational Nano-Materials Design on Green Energy'', ALCA of JST (``Spinodal Nanotechnology for Super-High-efficiency Energy Conversion''), the Global Center of Excellence (COE) Program, the ``Core Research and Engineering of Advanced Materials Interdisciplinary Education Center for Materials Science'', the Ministry of Education, Culture, Sports, Science and Technology, Japan, and a Grant in Aid for Scientific Research on Innovative Areas ``Materials Design through Computics: Complex Correlation and Non-Equilibrium Dynamics''.
One of Authors (HKY) thank to Prof. Hiroyuki Shiba (Tokyo Institute of Technology), Prof. Hisazumi Akai (University of Tokyo, ISSP) and Prof. Satoshi Fujimoto (Osaka University) for the valuable information and discussions about the computational materials design, and present status of the theory and quantum Monte Carlo simulation on negative $U_{\rm eff}$ Hubbard model, and also the relation to the BCS and BES.
HKY also thank to The Future Research Initiative Group Support Project on ``Computational Nano-Materials Design: New Strategic Materials'' in Osaka University.
\section*{References}
\bibliographystyle{unsrt}
\bibliography{refmain.bib}

\end{document}